\documentclass[fleqn]{elsart}
\usepackage{amsmath,graphicx,amssymb,oldlfont}
\journal{Physica D}
\begin{document}
% FORMATTING
%\baselineskip=21pt
\hsize=468pt
\oddsidemargin=-.1in
%MACRO DEFINITIONS
\def\bs{\boldsymbol}

\begin{frontmatter}

\title{Resonant triad dynamics in weakly damped Faraday waves with two-frequency forcing}
\author[NWU,LU]{J. Porter\corauthref{cor}},
\corauth[cor]{Corresponding author.  Address: Department of Applied Mathematics, University of Leeds, 
Leeds, LS2 9JT, UK.  Tel.: +44-113-343-5149; fax: +44-113-343-5090.}
\ead{jport@maths.leeds.ac.uk}
\author[NWU]{M. Silber}
\address[NWU]{Department of Engineering Sciences and Applied Mathematics,
Northwestern University, Evanston, Illinois 60208}
\address[LU]{Department of Applied Mathematics, University of Leeds, 
Leeds, LS2 9JT, UK}

\begin{abstract}
Many of the interesting patterns seen in recent multi-frequency Faraday experiments 
can be understood on the basis of three-wave interactions (resonant triads).  In this paper we 
consider two-frequency forcing and focus on a resonant triad that occurs near the 
bicritical point where two pattern-forming modes with distinct wavenumbers emerge 
simultaneously.   This triad has been observed directly (in the form of rhomboids) and has 
also been implicated in the formation of quasipatterns and superlattices.  We show 
how the symmetries of the undamped unforced problem (time translation, time reversal, 
and Hamiltonian structure) can be used, when the damping is weak, to obtain general 
scaling laws and additional qualitative properties of the normal form coefficients governing 
the pattern selection process near onset; such features help to explain why this particular 
triad is seen only for certain ``low" forcing ratios, and predict the existence of drifting 
solutions and heteroclinic cycles.  We confirm the anticipated parameter dependence 
of the coefficients and investigate its dynamical consequences using coefficients derived 
numerically from a quasipotential formulation of the Faraday problem due to Zhang and Vi\~nals.
\end{abstract}

\begin{keyword}
Resonant triads \sep Faraday waves \sep Pattern formation \sep Symmetry-breaking 
\PACS 05.45.-a \sep 47.20.Ky \sep 47.54.+r \sep 47.35.+i
\end{keyword}

\end{frontmatter}
%%%%%%%%%%%%%%%%%%%%%%%%%%%%%%%%%%%%%%%%%
%%%%%%%%%%%%%%%%%%%%%%%%%%%%%%%%%%%%%%%%%
\section{Introduction}
\label{sec:intro}

When a container of fluid is shaken vertically with sufficient strength, patterns of standing waves (SW) 
develop on the free surface.   This well-known transition, first described by Faraday \cite{Far1831}, is 
one of the best known examples of a pattern forming instability.  At a critical value of the forcing 
amplitude the flat surface loses stability and, with the aid of small perturbations, gives way to a spatially 
structured state.  This excited state is often highly ordered and has been observed 
in the form of rolls, squares, hexagons, and more exotic structures such as superlattices and 
quasi-patterns having 8, 10, or 12-fold rotational symmetry 
\cite{BinWat97,ArbFin02,EdwFau94,ArbFin98,KudPieGol98}.   The temporal spectrum of 
these resonant SW oscillations is highly ordered as well.   In the standard Faraday experiment with 
single frequency forcing of period $T=2\pi/\omega$ the first pattern to appear is a {\it subharmonic} 
response, i.e., it is $2T$-periodic in time \cite{BenUrs54}.  The characteristic wavenumber $k$ 
of the pattern is determined by the dispersion relation $\Omega(k)$ for gravity-capillary waves 
through the subharmonic resonance condition $\Omega(k)=\omega/2$.  This frequency $\omega$, 
which controls the spatial length scale, and the forcing amplitude $f$ are the most readily accessible 
control parameters.  

Recently, a number of investigators 
\cite{EdwFau94,ArbFin98,KudPieGol98,ArbFin00,PorSil02,TopSil02,SilSke99,Mul93} 
have looked at more general kinds of periodic forcing.   Such forcing functions have multiple 
frequency components and present a larger number of adjustable parameters.   One of the biggest 
advantages to using multiple frequencies is that one can study, in a controlled manner, the competition 
between modes of different length scales.   If the applied acceleration $f(t)$ is composed of two 
commensurate frequencies of ratio $m$:$n$ ($m$ and $n$ coprime):
\begin{align}
f(t) &= \frac{1}{2}\left(f_me^{im\omega t} + f_ne^{in\omega t} + c.c.\right) 
= |f_m|\cos(m\omega t + \phi_m) + |f_n|\cos(n\omega t + \phi_n),
\label{eq:forcing}
\end{align}   
an appropriate balance between $|f_m|$ and $|f_n|$ ensures that the two modes (one driven primarily  
by the $f_m$ forcing component and the other by $f_n$) onset simultaneously.  This {\it bicritical} 
point is the natural organizing center for the two-frequency problem because it is at this point that 
the greatest variety of patterns evolve and compete with each other while still in the weakly nonlinear 
regime.   A complete description of this competition is exceedingly difficult to achieve, yet a number 
of simple resonant interactions have emerged as crucial building blocks for many of the more 
interesting patterns.   In this respect resonant triads are particularly influential because they 
are the lowest order nonlinear effect (i.e., they arise at second order in a weakly nonlinear 
expansion) and provide a strong phase coupling between the three waves involved.   They have been 
implicated in a number of pattern formation studies, both experimental 
\cite{ArbFin02,EdwFau94,ArbFin98,ArbFin00,WagMulKno00} and  theoretical 
\cite{TopSil02,ZhaVin97_1,ZhaVin97_2,SilTopSke00,Mil91}.

\begin{figure}[ht]
\centerline{\includegraphics[width=3.6in]{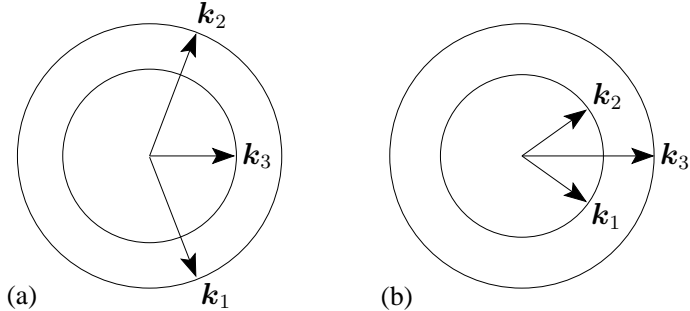}}
\caption{Resonant triad interactions at the bicritical point: (a) when $k_3<k_1$ and (b) when 
$k_1<k_3<2k_1$.}
\label{fig:critcirc}
\end{figure}

In this paper we consider the two-frequency forcing of Eq.~(\ref{eq:forcing}) and focus on a 
simple three-wave interaction, occurring near the bicritical point, in which two wavevectors 
from one critical circle, ${\bs k}_1$ and ${\bs k}_2$, sum to give a third wavevector ${\bs k}_3$ 
on the second critical circle (see Fig.~\ref{fig:critcirc}).  
%%%%%%%%%%%%%%%%%%%%%%%%%%%%%%%%%%%%%%%%FIG
%%%%%%%%%%%%%%%%%%%%%%%%%%%%%%%%%%%%%%%%FIG
Patterns formed by these three SW modes were called two-mode rhomboids by Arbell and 
Fineberg \cite{ArbFin00}.  The same basic resonance, however, was shown to be an essential component 
of more complicated states such as quasipatterns with 8-fold or 10-fold symmetry~\cite{ArbFin00} 
(the former observed with three-frequency forcing~\cite{ArbFin02}), and several other multiple mode 
and modulated states~\cite{ArbFin02}.   One of the most striking facts about this resonant triad is 
that it has so far been seen only with forcing ratios of 1:2 (as part of a superlattice-II type state \cite{ArbFin02}), 
2:3, and 4:5.  In \cite{PorSil02} we showed that the absence of two-mode rhomboids for larger 
values of $m$ and $n$ is related to the presence of (broken) time translation symmetry.   
This paper provides additional details behind that result as well as others in \cite{PorSil02}, and 
explores their dynamical implications.  

Symmetry arguments, in general, play a central role in our understanding of pattern formation, 
providing a powerful framework that explains generic features, i.e., properties that one expects 
to find across a wide range of physical systems sharing the same symmetry.   For Faraday systems of 
large aspect ratio (i.e., where the horizontal extent of the system is large compared to the 
characteristic wavelength) the relevant symmetry is that of the plane, described by the Euclidean 
group E(2).  If the pattern is spatially periodic, as is often the case near onset, 
many useful results (see, e.g., \cite{GolSteSha88,CraKno91}) can be obtained by restricting to patterns 
commensurate with this periodic lattice.  The choice of a lattice fixes the 
orientation and removes the arbitrary rotations that are part of E(2).   Spatial translations, 
however, are still allowed, as are certain discrete rotations ($60^\circ$ for hexagons, for 
example) and reflections.  Thus, when periodic planforms are considered 
the noncompact group E(2) is replaced by a compact group, namely, the semi-direct product of a 
discrete group representing the lattice symmetries (this group is $D_2$ for the resonant triads we 
consider) and the two-torus representing spatial translations.   The fact that these symmetry 
operations must transform a given solution into an equivalent solution restricts the form 
of the equations governing the pattern formation process near onset and thus determines which types 
of nonlinear terms can be present and which cannot \cite{GolSteSha88}.  

Temporal symmetries also play an important role in periodically forced systems.  In particular, such 
systems are invariant under discrete translation through one forcing period: $t \rightarrow t+T$.   Whether 
this symmetry is important or not depends on the nature of the linear eigenfunctions.  In the case of 
single-frequency forcing, for example, the eigenfunctions are subharmonic (i.e., they have odd 
parity under $t \rightarrow t+T$) and the associated ${\mathbb Z}_2$ symmetry bars all even terms from 
the evolution equations; in particular, the absence of quadratic terms means that there are no 
resonant triad interactions.  With two-frequency forcing the restrictions due to discrete temporal 
translation are less severe because the period $T$ is longer than the corresponding 
period for, say, isolated $m\omega$ forcing, which would be $T/m$.   In this way, modes driven 
``subharmonically'' by $f_m$ may actually be harmonic with respect to the full period $T$; 
they are harmonic if $m$ is even and subharmonic if $m$ is odd, similarly for $n$.

In this paper we follow the approach outlined in \cite{PorSil02} and extend the usual analysis 
based on ``exact'' spatial symmetries and discrete time translation to include the approximate 
(i.e., weakly broken) symmetries of continuous time translation, time reversal, and Hamiltonian 
structure.   In \cite{PorSil02} we considered hexagons and two-mode 
superlattices~\cite{ArbFin02,ArbFin98} as well as two-mode rhomboids, and showed 
how these symmetries can be used to determine the  leading order dependence of important 
{\it coefficients} in the SW amplitude equations.   In this paper we supply the details of that 
calculation for the case of two-mode rhomboids (here simply referred to as resonant triads) 
and explore the dynamical consequences of its predictions.   For example, in Section~\ref{sec:dyn} 
we show that a preference for resonant coefficients of opposite sign~\cite{PorSil02}, which stems 
from the presence of Hamiltonian structure in the undamped problem, leads to the existence of drifting 
solutions, modulated waves, and heteroclinic cycles.

The organization of the paper is as follows.  In Section~\ref{sec:sym} we demonstrate how the broken 
temporal symmetries can be replaced by unbroken {\it parameter} symmetries (see also \cite{PorSil02}), 
and use these in conjunction with the spatial symmetries to determine not only the form of the SW 
amplitude equations but also the manner in which the coefficients of these equations depend on the 
forcing parameters ($m$, $n$, $|f_m|$, $|f_n|$, $\phi_m$, $\phi_n$) and on the damping.   In 
Section~\ref{sec:ZV} we test our symmetry-based predictions against coefficients calculated 
numerically from a quasipotential formulation of the Faraday problem due to Zhang and 
Vi\~nals~\cite{ZhaVin97_1,ZhaVin97_2}.  Section~\ref{sec:dyn} summarizes the typical 
dynamics associated with the resonant triad equations and uses numerical examples to illustrate the 
effects of the broken symmetries discussed in Section~\ref{sec:sym}.  In 
Section~\ref{sec:conclusion} we offer our conclusions.

%%%%%%%%%%%%%%%%%%%%%%%%%%%%%%%%%%%%%%%%%
%%%%%%%%%%%%%%%%%%%%%%%%%%%%%%%%%%%%%%%%%
\section{Derivation of the Amplitude Equations}
\label{sec:sym}

In this section we use general symmetry arguments to derive the form of the equations governing 
resonant triads in the presence of two-frequency forcing (Eq.~(\ref{eq:forcing})).  In particular, we 
obtain general scaling properties of the normal form coefficients.  These scaling laws rest on the 
assumption of weak dissipation (the critical forcing amplitude needed to destabilize the spatially 
uniform state is then similarly small).  In this limit the problem is constrained not only by spatial 
symmetries but by the (weakly broken) symmetries of time translation $t \rightarrow t+s$, 
$s \in {\mathbb R}$ and time reversal $t \rightarrow -t$ that are features of the unforced 
inviscid hydrodynamical problem.  In fact, if these two symmetries are extended in an appropriate 
way (see Section~\ref{ssec:unfold}) they apply to the damped forced problem as well.  Additional 
constraints resulting from full Hamiltonian structure in the undamped forced problem are examined in 
Section~\ref{sssec:Ham}.   

There are many adjustable parameters in the two-frequency Faraday experiment.   In addition to the 
fluid depth (assumed to be infinite in our numerical study; see Section~\ref{sec:ZV}) and  
properties such as viscosity $\nu$, density $\rho$, and surface tension 
$\Gamma$, one can vary the basic frequency $\omega$ of the driving, the ratio of the two 
frequency components $m$:$n$, their respective magnitudes $|f_m|$ and $|f_n|$, and their 
phases $\phi_m$ and $\phi_n$.  Here we take $1/\omega$ as the unit of time, writing 
$\tau = \omega t$, and use the wavenumber $k_0$ associated with $\Omega_0 \equiv 
{\rm min}(m,n) \omega/2$ via the inviscid dispersion relation $g_0k_0 + \Gamma k_0^3/\rho = 
\Omega_0^2$ ($g_0$ is the acceleration of gravity) as the inverse unit of length.   This scaling 
groups $\omega$, $\nu$, $\Gamma$, and $\rho$ into three nondimensional parameters: 
$\gamma \equiv 2\nu k_0^2/\omega$ (the damping), $\Gamma_0 \equiv \Gamma k_0^3/(\rho \omega^2)$ 
(the capillarity number), and $G_0 \equiv g_0k_0/\omega^2$ (the gravity number).  The last two are 
related by the equation $\Gamma_0 + G_0 = ({\rm min}(m,n)/2)^2$.   

\begin{figure}[ht]
\centerline{\includegraphics[width=3.2in]{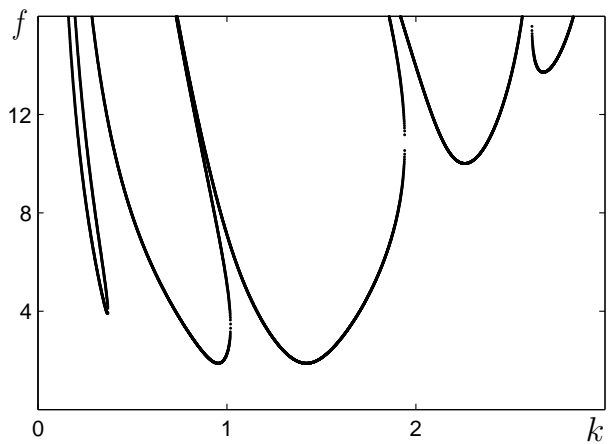}}
\caption{Neutral stability curves at the bicritical point calculated from the Zhang-Vi\~nals equations 
(see Section~\ref{sec:ZV}).  We use $m$:$n =3$:2, $\gamma = 0.1$, $\Gamma_0 = 0.5$, $G_0 = 0.5$, 
$\phi_m=\phi_n=0$ and $|f_2|/|f_3| \simeq 0.423$.   The applied acceleration $f$ is in units of $g_0$.}
\label{fig:23neut}
\end{figure}

The situation of special interest here is the bicritical point where two modes of distinct wavenumber 
are simultaneously excited as the overall forcing amplitude $f \equiv (|f_m|^2+|f_n|^2)^{1/2}$ is 
increased (see Fig.~\ref{fig:23neut}).  
%%%%%%%%%%%%%%%%%%%%%%%%%%%%%%%%%%%%%%%%FIG
%%%%%%%%%%%%%%%%%%%%%%%%%%%%%%%%%%%%%%%%FIG
We think of this bistability criteria as determining the ratio $|f_m|/|f_n|$.  

The two critical circles at the bicritical point and the two different ways of constructing a resonant 
triad are illustrated in Fig.~\ref{fig:critcirc}.  In either case we choose  three wavevectors: two of equal 
magnitude, $|{\bs k}_1| = |{\bs k}_2| = k_1$, and their sum, ${\bs k}_3 = {\bs k}_1 + {\bs k}_2 $.  
Without loss of generality we associate ${\bs k}_1$ and ${\bs k}_2$ with the $f_m$ forcing 
component and ${\bs k}_3$ with the $f_n$ component.  Thus Fig.~\ref{fig:critcirc}a applies if 
$m > n$ while Fig.~\ref{fig:critcirc}b applies if $m < n$.   As mentioned, when the damping is 
small each of the excited modes can, to first approximation, be thought of as a subharmonic response 
to the corresponding component ($f_m$ or $f_n$) of the driving.   The $\exp\{i{\bs k}_1 \cdot {\bs x}\}$ 
and $\exp\{i{\bs k}_2 \cdot {\bs x}\}$ modes oscillate with a dominant frequency of $m/2$ while the 
$\exp\{i{\bs k}_3 \cdot {\bs x}\}$ mode has dominant frequency $n/2$.   A given mode is 
harmonic (subharmonic) when $m$ or $n$ is even (odd).  As shown in Section~\ref{ssec:unfold}, 
resonant triads only occur if $n$ is even, in which case $m$ is odd since $m$ and $n$ are coprime.

As long as damping is present the first modes to appear as $f$ is increased are SW.  Sufficiently 
close to onset the resonant triad system may therefore be described by the three complex amplitudes 
$A_j$, $j=1,2,3$ of the SW modes $\exp\{i \bs{k}_j \cdot \bs{x}\}$.   We do not begin immediately 
with these three SW modes but, following the program of \cite{PorSil02}, obtain first the form of the 
equations governing six traveling wave (TW) modes having the same wavevectors ${\bs k}_1$, 
${\bs k}_2$, and ${\bs k}_3$.   It should be noted that in the absence of dissipation and forcing the 
Faraday system supports TW at infinitely many frequencies and wavenumbers, i.e., the only 
requirement is from the inviscid dispersion relation $\Omega(k)$.  It is precisely the applied 
forcing which selects, through the temporal resonance condition, the frequencies and hence the 
length scales of the problem.   Our approach, by restricting attention to the wavenumbers $k_1$ and 
$k_3$, anticipates the effect of the parametric forcing.  At the same time, however, it assumes that the 
amplitude of this forcing (and of the damping) is small.  In this limit a TW expansion is justified despite 
the fact that pure TW solutions are destroyed by the parametric forcing term -- such an approach may 
also be viewed as the unfolding of a Hopf bifurcation (occurring at $\gamma=0$) with nearly resonant 
temporal forcing (see \cite{RieCraKno88,RieSilKra94}).   Here we are motivated in our use of TW 
by the action of the weakly broken temporal symmetries; TW transform in a simple way under time translation 
$\tau \rightarrow \tau+s$ and time reversal $\tau \rightarrow -\tau$ whereas SW do not.   The simple 
transformation properties of TW allow us to take full advantage of the spatial and temporal symmetries 
before a center manifold reduction is used to obtain the equations describing the three SW amplitudes $A_j$.   

We now systematically consider the two unfolding parameters which move the system away from 
the ``ideal'' undamped, unforced case: the damping parameter $\gamma >0$ (proportional to viscosity) that  
breaks time reversal symmetry, and the forcing function $f(\tau)$ (equivalently, $f_m$ and $f_n$), 
which breaks the continuous time translation symmetry and, in general, the time reversal symmetry as 
well.   In the process we pay special attention to resonant terms, i.e., those terms which couple the phases 
of the different modes, because of the crucial role they play in the ensuing dynamics.  The leading resonant 
terms are quadratic in amplitude but, with the exception of $m$:$n$ = 1:2, appear only 
as a result of broken time translation symmetry.   The residual effect of this symmetry determines, 
for small $\gamma$, the influence (i.e., size) of these resonant terms by prescribing how they 
scale with the symmetry-breaking parameters $f_m$ and $f_n$.

%%%%%%%%%%%%%%%%%%%%%%%%%%%%%%%%%%%%
\subsection{Traveling wave equations}

We consider a system of six undamped, unforced TW with wavevectors ${\bs k}_1$, ${\bs k}_2$, 
and ${\bs k}_3$.  For example, the surface height $h({\bs x},\tau)$ is written as
\begin{align}
h({\bs x},\tau) = &\,Z_1^+e^{i({\bs k}_1\cdot {\bs x}+m  \tau/2)} + 
Z_1^-e^{i({\bs k}_1\cdot {\bs x}- m \tau/2)} + Z_2^+e^{i({\bs k}_2\cdot{\bs x}+ m  \tau/2)} + 
Z_2^-e^{i({\bs k}_2\cdot {\bs x}- m \tau/2)}\nonumber \\ 
&\, Z_3^+e^{i({\bs k}_3\cdot {\bs x}+ n  \tau/2)} + 
Z_3^-e^{i({\bs k}_3\cdot {\bs x}- n  \tau/2)} + c.c. + ...\,.
\label{eq:TWexpansion}
\end{align} 
The equations describing the slow evolution of the TW amplitudes $Z_j^\pm$, $j=1,2,3$  
must be equivariant under the operations of spatial translation (${\cal T}_{\bs \phi}$), 
reflection ($\kappa$) about a line through ${\bs k}_3$, and rotation ({$\cal R$}) by $\pi$ (i.e., 
inversion through the origin):
\begin{align}
{\cal T}_{\bs \phi}&: Z_j^\pm \rightarrow Z_j^\pm e^{i\phi_j},
\quad {\bs \phi}=(\phi_1,\phi_2), \quad \phi_1, 
\phi_2 \in [0,2\pi), \quad \phi_3=\phi_1+\phi_2, \label{sym1:T}\\
\kappa &:  Z_1^\pm \leftrightarrow Z_2^\pm,\label{sym1:kappa}\\
{\cal R}&: Z_j^\pm \rightarrow \bar{Z}_j^\mp.\label{sym1:R}
\end{align}  %]
In the absence of forcing ($f$) the equations must also be equivariant under time translation:
\begin{alignat}{2}
{\cal T}_s : (Z_1^\pm,Z_2^\pm) &\rightarrow (Z_1^\pm,Z_2^\pm)e^{\pm im s/2},\qquad 
&Z_3^\pm &\rightarrow Z_3^\pm e^{\pm ins/2},\qquad s \in [0,4\pi),
\label{sym:timetrans}
\end{alignat}  %]
while if both $f$ and $\gamma$ are zero there is the time reversal symmetry $\sigma$:
\begin{align}
\sigma \,:\, \tau \rightarrow -\tau, \quad Z_j^\pm \rightarrow Z_j^\mp,  
\label{sym:timerev}
\end{align}  
The constraints of symmetries~(\ref{sym1:T}-\ref{sym:timerev}) lead to equations, truncated 
at cubic order, of the form
\begin{align}
\dot{Z}_1^+ =\,&\{\tilde{\alpha}\bar{Z}_2^+ Z_3^+ \}_{1:2} + 
\delta Z_2^+ \bar{Z}_2^- Z_1^- \nonumber \\  
+ &\left({\rm a}|Z_1^+|^2 + {\rm b}|Z_1^-|^2 + {\rm c}|Z_2^+|^2 + {\rm d}|Z_2^-|^2 + 
{\rm e}|Z_3^+|^2 + {\rm f}|Z_3^-|^2 \right)Z_1^+,
\label{eq:TW1}\\
\dot{Z}_3^+ =\,&\{\tilde{\beta}Z_1^+ Z_2^+\}_{1:2} + \left({\rm g}(|Z_1^+|^2 + |Z_2^+|^2) 
+ {\rm h}(|Z_1^-|^2 + |Z_2^-|^2) + {\rm l}|Z_3^+|^2 + {\rm p}|Z_3^-|^2  \right)Z_3^+,
\label{eq:TW3} 
\end{align}
where the overdot indicates a derivative with respect to $\tau$.  The other four equations follow from 
the symmetries $\kappa$ and ${\cal R}$.  The quadratic terms in curly brackets (subscript 1:2) 
are permitted only when $m$:$n$ = 1:2 because with other forcing ratios they violate the time 
translation symmetry ${\cal T}_s$.  Whether these resonant terms are present or not, the time 
reversal symmetry $\sigma$ (more precisely, $\sigma \circ {\cal R}$) forces all coefficients to be 
purely imaginary: 
\begin{align}
\tilde{\alpha} = i\tilde{\alpha}_i, \quad \delta = i\delta_i,
\quad a=ia_i,\quad ...,\quad p=ip_i.
\end{align} 
The absence of linear terms with imaginary coefficients is not related to symmetry but to 
the way we define $k_1$ and $k_3$, which requires that the detuning vanish at $f=\gamma=0$.  
To see this, recall that expansion~(\ref{eq:TWexpansion}) anticipates finite damping and forcing in two 
ways: it assumes the TW modes are in subharmonic resonance with the two driving frequencies, i.e., they 
oscillate at  $m/2$ or $n/2$, and it takes the wavenumbers $k_1$ and $k_3$ from the $\gamma$-dependent 
dual minima of the SW neutral stability curves at the bicritical point of the damped forced problem 
{\it not} from the dispersion relation, $\Omega(k_1)=m/2$ and $\Omega(k_3)=n/2$.  One 
consequence of this implicit $\gamma$ dependence in $k_1$ and $k_3$ is a detuning in the damped 
forced problem between the parametrically excited frequencies $m/2$ and $n/2$ and the ``natural'' 
frequencies $\Omega(k_1)$ and $\Omega(k_3)$.  In Eqs.~(\ref{eq:TW1},\ref{eq:TW3}) however, 
the damping and forcing are both zero, and the wavenumbers $k_1$ and $k_3$ coincide with those 
given by the inviscid dispersion relation, i.e., $\Omega(k_1)=m/2$ and $\Omega(k_3)=n/2$.  
Consequently, there is no detuning at $f=\gamma=0$.

%%%%%%%%%%%%%%%%%%%%%%%%%%%%%%%%%%%%
\subsection{Unfolding of Eqs.~(\ref{eq:TW1}-\ref{eq:TW3}): parameter symmetries}
\label{ssec:unfold}

We now unfold the vector field~(\ref{eq:TW1},\ref{eq:TW3}) assuming all symmetry-breaking 
parameters are the same order: $|f_m| \sim |f_n| \sim \gamma \sim \epsilon$.   Throughout this 
unfolding we will be interested only in those terms which add new qualitative effects or 
determine the leading order scaling of coefficients in the subsequent SW equations.  Assuming 
that the symmetry-breaking terms are analytic in $f$ and $\gamma$, their precise form can be 
obtained by appealing to a more general class of symmetries in which parameters may be 
transformed along with dynamical variables (see, e.g., \cite{Swi88}).  In this way the broken 
symmetries ${\cal T}_s$ and $\sigma$ can be recast in terms of intact {\it parameter symmetries}.   Put 
another way, it is still possible to translate or reflect the problem in time so long as the damping and 
forcing terms are also appropriately transformed.  Specifically, we have the parameter symmetry 
\begin{alignat}{2}
{\cal T}_s^{\rm par} :\, &(Z_1^\pm,Z_2^\pm) \rightarrow (Z_1^\pm,Z_2^\pm)e^{\pm ims/2}, \qquad 
&Z_3^\pm \rightarrow Z_3^\pm e^{\pm ins/2},\nonumber \\
&(f_m,f_n) \rightarrow (f_m e^{i m s}, f_n e^{i n s}),
\label{sym:partime}
\end{alignat}
replacing ${\cal T}_s$, while $\sigma$ becomes
\begin{align}
\sigma^{\rm par} \,:\, (\tau,\gamma) \rightarrow -(\tau,\gamma), 
\quad Z_j^\pm \rightarrow Z_j^\mp, \quad (f_m,f_n) \rightarrow (\bar{f}_m,\bar{f}_n).
\end{align}
Although parameter symmetries are sometimes unphysical (it makes little sense to take 
$\gamma < 0$) they are perfectly good mathematical symmetries of the governing 
equations and may be used to advantage - in this case they yield the proper form of 
symmetry-breaking terms.

%%%%%%%%%%%%%%%%%%%%%%%%%%%%%%%%%%%%
\subsubsection{Finite forcing: $f \neq 0$}

The presence of forcing allows several new terms in Eqs.~(\ref{eq:TW1},\ref{eq:TW3}).  
For example, at linear order in amplitude Eq.~(\ref{eq:TW1}) contains the usual parametric 
forcing term $f_m Z_1^-$, consistent with the symmetry ${\cal T}_s^{\rm par}$.   Another 
important addition comes in the form of quadratic resonance terms.  The quadratic amplitude 
dependence follows from the spatial resonance condition~(\ref{sym1:T}) but the dependence 
on $f_m$ and $f_n$ can only be determined with the help of ${\cal T}_s^{\rm par}$, i.e., it 
reflects a temporal resonance requirement.  To determine this dependence we consider adding 
terms, consistent with the spatial resonance requirement, of the form 
$f_m^r f_n^u \bar{Z}_2^\pm Z_3^\pm$ and $f_m^v f_n^w Z_1^\pm Z_2^\pm$ to 
Eqs.~(\ref{eq:TW1}) and (\ref{eq:TW3}), respectively.  Here $r$ ,$u$, $v$, and $w$ are integers, 
allowed to be negative to indicate complex conjugation (i.e., $f_m^{-2} \equiv \bar{f}_m^2$ and 
so on).   A combination such as $f_m\bar{f}_m$ is invariant under ${\cal T}_s^{\rm par}$ and appears 
only at higher order.  The temporal resonance condition, which follows from the equivariance 
requirement under ${\cal T}_s^{\rm par}$, may be expressed as 
\begin{align}
m/2 &= r\,m +  u\,n - \varepsilon_1 m/2 + \varepsilon_2 n/2,
\label{eq:res1}\\
n/2 &= v\,m + w\, n + \varepsilon_3 m/2 + \varepsilon_4 m/2,
\label{eq:res2}
\end{align}
where $\varepsilon_i = \pm 1$, $i=1,...,4$.  These $\varepsilon_i$ distinguish between 
the $Z_j^+$ modes (the corresponding $\varepsilon_i$ are positive) and the 
$Z_j^-$ modes (negative $\varepsilon_i$).  We seek values for the parameters $r$, $u$, $v$, $w$, 
$\varepsilon_1$, $\varepsilon_2$, $\varepsilon_3$, and $\varepsilon_4$ that minimize 
both $|r|+ |u|$ and $|v| + |w|$ and therefore give resonance terms of lowest order in $f$.  To 
accomplish this we rewrite Eqs.~(\ref{eq:res1}) and (\ref{eq:res2}):
\begin{align}
m(1+\varepsilon_1-2 r) &= n(\varepsilon_2+2 u),\label{eq:res3}\\
m(1+\varepsilon_3\varepsilon_4+2\varepsilon_3v) &= n(\varepsilon_3 
- 2\varepsilon_3 w),\label{eq:res4}
\end{align}
showing that, as claimed earlier, $n$ must be even else there is no solution at all, and 
that Eq.~(\ref{eq:res4}) is equivalent to Eq.~(\ref{eq:res3}) under $v \rightarrow -\varepsilon_2 r$, 
$w \rightarrow -\varepsilon_2 u$, $\varepsilon_3  \rightarrow \varepsilon_2$, and $\varepsilon_4  
\rightarrow \varepsilon_1\varepsilon_2$.  We may therefore focus on Eq.~(\ref{eq:res3}), which 
requires that the quantity $1+\varepsilon_1-2 r$ (hereafter $\Lambda$) is a nonzero multiple of n.  
In terms of $\Lambda$ the minimum of $|r|+|u|$ may be expressed (using Eq.~(\ref{eq:res3})) as the 
minimum of $G(\Lambda,\varepsilon_1,\varepsilon_2)\equiv|\Lambda-1-\varepsilon_1| 
+ |m\Lambda/n - \varepsilon_2|$.  For positive $\Lambda$ this minimum occurs when 
$\varepsilon_1=\varepsilon_2=1$ and yields $G(\Lambda,1,1) = (1+m/n)\Lambda - 3$ (the 
absolute values can be removed because $\Lambda \geq n \geq 2$).  Alternatively, if $\Lambda<0$ 
the minimum of $G(\Lambda,\varepsilon_1,\varepsilon_2)$ is realized with 
$\varepsilon_1=\varepsilon_2=-1$ and leads to $G(\Lambda,-1,-1) =(1+m/n)|\Lambda| - 1$.  
It follows that the global minima occurs for the smallest positive $\Lambda$, i.e., $\Lambda=n$, a 
choice which implies $r=-(n-2)/2$ and $u=(m-1)/2$.  The order of the resonance terms (in $f$) is 
thus $|r|+|u|=(n+m-3)/2$. 

Altogether, we may rewrite Eqs.~(\ref{eq:TW1},\ref{eq:TW3}) in the presence of forcing as
\begin{align}
\dot{Z}_1^+ =\; & \upsilon Z_1^+ - (\tilde{\lambda}f_m + \varpi \bar{f}_m^{n-1} f_n^m)Z_1^- 
+ \tilde{\alpha} \bar{f}_m^{\frac{n-2}{2}}f_n^{\frac{m-1}{2}} \bar{Z}_2^+ Z_3^+ + 
\delta Z_2^+\bar{Z}_2^- Z_1^- \nonumber \\
+\, &\left({\rm a}|Z_1^+|^2 + {\rm b}|Z_1^-|^2 + {\rm c}|Z_2^+|^2 + 
{\rm d}|Z_2^-|^2 + {\rm e}|Z_3^+|^2 + {\rm f}|Z_3^-|^2  \right)Z_1^+ \nonumber \\
+\,&\left({\rm q}_1|Z_1^+|^2 + {\rm q}_2|Z_1^-|^2 + {\rm q}_3|Z_2^+|^2 + 
{\rm q}_4|Z_2^-|^2 + {\rm q}_5|Z_3^+|^2 + {\rm q}_6|Z_3^-|^2  \right)f_m Z_1^- 
\nonumber \\ +\,&\left({\rm r}_1f_m \bar{Z}_2^+ Z_2^- + 
\bar{f}_m({\rm r}_2 Z_1^+\bar{Z}_1^- +{\rm r}_3 Z_2^+\bar{Z}_2^-) + 
{\rm r}_4f_n\bar{Z}_3^+ Z_3^- + {\rm r}_5 \bar{f}_n Z_3^+\bar{Z}_3^-\right)Z_1^+,
\label{eq:TWf1} \\
\dot{Z}_3^+=\; &\varrho  Z_3^+ - (\tilde{\mu} f_n + \zeta f_m^n\bar{f}_n^{m-1})Z_3^- 
+ \tilde{\beta} f_m^{\frac{n-2}{2}}\bar{f}_n^{\frac{m-1}{2}}Z_1^+ Z_2^+ \nonumber \\
+\,&\left({\rm g}(|Z_1^+|^2 + |Z_2^+|^2) + {\rm h}(|Z_1^-|^2 + |Z_2^-|^2) + 
{\rm l}|Z_3^+|^2 + {\rm p}|Z_3^-|^2  \right)Z_3^+ \nonumber \\
+\,&\left({\rm s}_1(|Z_1^+|^2 + |Z_2^+|^2) + {\rm s}_2(|Z_1^-|^2 + |Z_2^-|^2) 
+ {\rm s}_3|Z_3^+|^2 + {\rm s}_4|Z_3^-|^2  \right)f_n Z_3^- 
\nonumber \\ +\,&\left({\rm u}_1f_m(\bar{Z}_1^+Z_1^- +\bar{Z}_2^+Z_2^-) + 
{\rm u}_2\bar{f}_m(Z_1^+\bar{Z}_1^- +Z_2^+\bar{Z}_2^-) +
{\rm u}_3 \bar{f}_nZ_3^+\bar{Z}_3^-\right)Z_3^+.
\label{eq:TWf3}
\end{align}
The remaining equations follow from the symmetries $\kappa$ and ${\cal R}$.  At second and third  
order in the amplitudes $Z_j^\pm$ we have kept all terms which contribute at leading order 
(in $\epsilon$) to the corresponding SW coefficients (Eqs.~(\ref{eq:SWcoef}) of 
Section~\ref{ssec:reductiontoSW}).  At the linear level, however, we have included two 
additional terms (with coefficients $\varpi$ and $\varrho$) that do not contribute at leading 
order but are nonetheless responsible for an important qualitative effect, namely, a dependence in the 
linear problem on the relative phase of the two forcing components - this dependence will 
be discussed in more detail below.  We note that terms of the form $f Z^3$, for example 
${\rm q}_1|Z_1^+|^2f_mZ_1^-$ in Eq.~(\ref{eq:TWf1}), were also found by Zhang~\cite{Zha94}.

Because of time reflection $\sigma^{\rm par}$ most coefficients in 
Eqs.~(\ref{eq:TWf1},\ref{eq:TWf3}), as in Eqs.~(\ref{eq:TW1},\ref{eq:TW3}), are 
imaginary at leading order. They do, however, include small $f$-dependent corrections, 
invariant under ${\cal T}_s^{\rm par}$, which can generate real parts.  For example, we have
\begin{align*}
{\rm a} =  i({\rm a}_i+{\rm a}_m|f_m|^2+{\rm a}_n|f_n|^2 ) + 
{\rm a}_f {\rm Im}(f_m^n\bar{f}_n^m) + ...,
\end{align*}
where a$_i$, a$_m$, a$_n$, a$_f \in {\mathbb R}$.  Analogous expansions apply to all remaining 
coefficients except $\upsilon$ and $\varrho$ which contain no ${\cal O}(1)$ imaginary part 
(no detuning at $\gamma=f=0$).  Most $f$-dependent contributions to the coefficients of 
Eqs.~(\ref{eq:TWf1},\ref{eq:TWf3}) are unimportant.  In fact, with the exception of those 
in $\upsilon$, $\varrho$, $\tilde{\lambda}$ and $\tilde{\mu}$ (see below) we neglect all of them, 
keeping only the ${\cal O}(1)$ imaginary part and dropping the real parts proportional to 
${\rm Im}(f_m^n\bar{f}_n^m)$ in anticipation of lower order contributions due to finite 
$\gamma$.

%%%%%%%%%%%%%%%%%%%%%%%%%%%%%%%%%%%%
\subsubsection{Finite damping: $\gamma \neq 0$}

The principal effect of the damping $\gamma$ is to give real parts to all coefficients 
in Eqs.~(\ref{eq:TWf1},\ref{eq:TWf3}).  Due to $\sigma^{\rm par}$ these contributions 
must be odd under $\gamma \rightarrow -\gamma$ and take the form $\gamma$, $\gamma^3$, 
$\gamma |f_m|^2$, etc.  Additional corrections to the imaginary parts, which must be 
even under $\gamma \rightarrow -\gamma$, are functions of $\gamma^2$ and appear at 
order $\epsilon^2$ and higher.  Such imaginary corrections only matter for the detunings 
${\rm Im}(\upsilon_1)$ and ${\rm Im}(\upsilon_3)$, where they contribute at leading order 
to the SW coefficients, and in the (linear) parametric forcing coefficients $\tilde{\lambda}$ 
and $\tilde{\mu}$ where they contribute to a (phase-independent) coupling in the SW linear 
problem between the two forcing components $f_m$ and $f_n$ -- some of the 
${\cal O}(\epsilon^3)$ terms in the real parts of $\upsilon$ and $\varrho$ 
contribute to this coupling as well.  Put simply, TW of type $Z_{1,2}^\pm$ ($Z_3^\pm$) are 
primarily excited by $f_m$ ($f_n$) but are affected by $f_n$ ($f_m$) as well; we are 
interested in that effect.  For the four linear coefficients mentioned above we therefore write
\begin{align}
&\upsilon = \gamma (-\upsilon_r + \hat{\upsilon}_\gamma \gamma^2 + \hat{\upsilon}_m|f_m|^2 + 
\hat{\upsilon}_n|f_n|^2) + i(\upsilon_\gamma \gamma^2 + \upsilon_m|f_m|^2 + \upsilon_n|f_n|^2), 
\quad \upsilon_r >0,
\label{eq:upsilon}\\
&\varrho = \gamma (-\varrho_r + \hat{\varrho}_\gamma \gamma^2 + \hat{\varrho}_m|f_m|^2 + 
\hat{\varrho}_n|f_n|^2)+ i(\varrho_\gamma \gamma^2 + \varrho_m|f_m|^2+\varrho_n|f_n|^2), 
\quad\, \varrho_r > 0,
\label{eq:varrho}\\
&\tilde{\lambda} = \tilde{\lambda}_r\gamma +i(\tilde{\lambda}_i+\tilde{\lambda}_\gamma \gamma^2 
+\tilde{\lambda}_m|f_m|^2 + \tilde{\lambda}_n|f_n|^2),
\label{eq:lambda}\\
&\tilde{\mu} = \tilde{\mu}_r\gamma +i(\tilde{\mu}_i+\tilde{\mu}_\gamma \gamma^2 + 
\tilde{\mu}_m|f_m|^2+\tilde{\mu}_n|f_n|^2), 
\label{eq:mu}
\end{align}
while for the remaining coefficients we use the expansions
\begin{align}
&\varpi = \varpi_r \gamma + i \varpi_i, \qquad
\tilde{\alpha} = \tilde{\alpha}_r \gamma + i \tilde{\alpha}_i, \qquad 
\delta = \delta_r \gamma + i\delta_i, \qquad {\rm etc.},
\label{eq:coef2}
\end{align}
stopping at lowest order in both real and imaginary parts.   

The form of expansions~(\ref{eq:upsilon}-\ref{eq:coef2}) is consistent (where it overlaps) 
with the analytical formulas derived by Topaz and Silber~\cite{TopSil02} in the case of 
one-dimensional waves.  Those results were also for two-frequency forcing in the case of weak 
damping, and were based on the Zhang-Vi\~nals equations.   One difference is that 
Topaz and Silber find no ${\cal O}(\epsilon)$ real part to $\tilde{\lambda}$, that is, 
$\tilde{\lambda}_r = 0$ (one would likewise expect $\tilde{\mu}_r = 0$).  Setting 
$\tilde{\lambda}_r = \tilde{\mu}_r = 0$ is of no consequence for us, however, because those  
terms do not contribute at leading order to the SW coefficients (see Eqs.~(\ref{eq:SWcoef})).

It is of interest now to consider the linear problem and, in particular, the effect 
(mentioned above) of the phases ($\phi_m$ and $\phi_n$ in Eq.~(\ref{eq:forcing})) on 
the bicritical point defined by
\begin{alignat}{2}
&|\upsilon|=|F_m|,\qquad  &&F_m \equiv \tilde{\lambda}f_m + 
\varpi\bar{f}_m^{n-1}f_n^m, 
\label{eq:fmc}\\
&|\varrho|=|F_n|,\qquad  &&F_n \equiv \tilde{\mu}f_n 
+\zeta f_m^n\bar{f}_n^{m-1}.
\label{eq:fnc}
\end{alignat}
At leading order this bicritical point is independent of $\phi_m$ and $\phi_n$, i.e., the 
critical amplitudes $|f_m^{\rm c}|$ and $|f_n^{\rm c}|$ do not couple and we have simply
\begin{align}
&|f_m^{\rm c}| = \frac{\upsilon_r}{|\tilde{\lambda}_i|}\gamma, 
\qquad |f_n^{\rm c}| = \frac{\varrho_r}{|\tilde{\mu}_i|}\gamma. 
\label{eq:fc1}
\end{align}
Expressions~(\ref{eq:fc1}) are utilized throughout this paper (for example, in evaluating 
coefficients~(\ref{eq:upsilon}-\ref{eq:mu}) at the bicritical point to obtain 
expressions~(\ref{eq:J1}-\ref{eq:J7}) of the Appendix) and are sufficient to obtain 
the leading order terms in the SW normal form coefficients (see Eqs.~(\ref{eq:SWcoef})).   
However, the lack of phase dependence in Eqs.~(\ref{eq:fc1}) is contrary to the experimental 
observations of M\"uller~\cite{Mul93} and the calculations of Zhang and Vi\~nals \cite{ZhaVin97_2} 
for the case $m$:$n = 1$:2.  These authors found that there was a strong dependence of 
$r \equiv |f_m^{\rm c}|/(|f_m^{\rm c}|+|f_n^{\rm c}|)$ on $\phi_n$.  To see this dependence 
Eqs.~(\ref{eq:fmc},\ref{eq:fnc}) must be solved to high enough order to capture the 
influence of the $\varpi\bar{f}_m^{n-1}f_n^m$ and $\zeta f_m^n\bar{f}_n^{m-1}$ terms.  
When this is done for $m$:$n = 1$:2 the expression for $r$ is
\begin{align}
r = \frac{\upsilon_r|\tilde{\mu}_i|}{\upsilon_r|\tilde{\mu}_i|+\varrho_r|\tilde{\lambda}_i|}
+ \frac{\zeta_i\tilde{\mu}_i\upsilon_r^3-\varpi_i\tilde{\lambda}_i\upsilon_r\varrho_r^2}
{|\tilde{\lambda}_i|(\upsilon_r|\tilde{\mu}_i|+\varrho_r|\tilde{\lambda}_i|)^2}\gamma\cos\phi_f,
\label{eq:fcofphi}
\end{align}
which reveals an ${\cal O}(\epsilon)$ harmonic dependence on the relative phase
\begin{align}
\phi_{f}=m\phi_n-n\phi_m.
\label{eq:phif}
\end{align}
This particular linear combination of $\phi_m$ and $\phi_n$, the complex phase of the 
${\cal T}_s^{\rm par}$-invariant term $f_n^m\bar{f}_m^n$, is the only physically relevant one.  
In general, the $\phi_f$-dependent correction to $r$ is ${\cal O}(\epsilon^{m+n-2})$ 
and is much less important (with small damping) when $m+n$ is large.  For the  
next available resonant triad interaction, $m$:$n = 3$:2, it is already of third order.  
The diminishing influence of $\phi_f$ with $m+n$ is consistent with the experiments of 
Edwards and Fauve \cite{EdwFau94} where no significant phase dependence was found for 
$m$:$n = 5$:4.  In the case $m$:$n = 1$:2 we can compare Eq.~(\ref{eq:fcofphi}) with 
\cite{ZhaVin97_2}, where a different convention for the two-frequency forcing function is used, by setting 
$\phi_m = \pi/2$ and $\phi_n = \pi/2 + \phi$ ($\phi$ is the phase variable used in \cite{ZhaVin97_2}) 
to find $\phi_f = \phi - \pi/2$.  Thus Eq.~(\ref{eq:fcofphi}) predicts an extremum for $r(\phi)$ 
at $\phi = \pi/2$, in agreement with the results of \cite{Mul93} and \cite{ZhaVin97_2} (see Fig.~1 of 
\cite{ZhaVin97_2}) -- to be more precise, a maximum was observed at $\phi = \pi/2$.  That a 
maximum should be observed rather than a minimum cannot be determined from 
Eq.~(\ref{eq:fcofphi}) since this depends on the actual numerical values of the coefficients 
$\upsilon_r$, $\tilde{\mu}_i$, etc..

%%%%%%%%%%%%%%%%%%%%%%%%%%%%%%%%%%%%
\subsection{Reduction to standing wave equations}
\label{ssec:reductiontoSW}

The ${\cal O}(\epsilon)$ linear coupling between $Z_j^+$ and $Z_j^-$ due to the parametric 
forcing destroys the pure TW solutions of Eqs.~(\ref{eq:TW1},\ref{eq:TW3}) and ensures that 
the first instability with increasing $f$ produces SW satisfying $|Z_j^+|=|Z_j^-|$.  A standard 
reduction procedure (see, for example, \cite{Cra91}) may therefore be used to reduce 
Eqs.~(\ref{eq:TWf1},\ref{eq:TWf3}), near the bicritical point, to a system of three amplitude 
equations describing SW modes.  Small deviations from the bicritical point are captured with the 
reduced forcing parameters $\Delta_m, \Delta_n \in {\mathbb R}$ according to 
$f_m = f_m^{\rm c}(1+\Delta_m)$,  $f_n = f_n^{\rm c}(1+\Delta_n)$.  To facilitate the reduction we 
first rescale the amplitudes so that (after appropriate redefinition) the complex phases of 
$F_m$ and $F_n$ are equal, respectively, to those of $\upsilon$ and $\varrho$.   
Specifically, this rescaling takes $(Z_1^\pm,Z_2^\pm) \rightarrow (Z_1^\pm,Z_2^\pm)
\exp\{\pm i\vartheta_1/2\}$ and $Z_3^\pm \rightarrow Z_3^\pm \exp\{\pm i\vartheta_3/2\}$, where 
$F_m^{\rm c} \bar{\upsilon}^{\rm c} = |F_m^{\rm c} \upsilon^{\rm c}| 
\exp\{i\vartheta_1\}$ and $F_n^{\rm c} \bar{\varrho}^{\,\rm c} = |F_n^{\rm c} 
\varrho^{\,\!\rm c}| \exp\{i\vartheta_3\}$.   In addition to $F_m$ and $F_n$, 
phase shifts are induced in all of the nonlinear coefficients in Eqs.~(\ref{eq:TWf1},\ref{eq:TWf3}) 
except a--h, l, and p.   As a result of this rescaling the critical eigenmodes satisfy the simple 
relation $Z_j^+=Z_j^- \equiv A_j$.  

After the reduction we find that the $A_j$ satisfy
\begin{align}
\dot{A}_1 &= \lambda A_1 + \alpha \bar{A}_2 A_3 + 
A_1\left(a|A_1|^2 + b|A_2|^2+c|A_3|^2\right),
\label{eq:SW1}\\
\dot{A}_2 &= \lambda A_2 + \alpha \bar{A}_1A_3 + 
A_2\left(a|A_2|^2 + b|A_1|^2+c|A_3|^2\right),
\label{eq:SW2}\\
\dot{A}_3 &= \mu A_3 + \beta A_1 A_2 + A_3\left(d|A_1|^2 + 
d|A_2|^2+e|A_3|^2\right),
\label{eq:SW3}
\end{align}
where 
\begin{align}
&\lambda = \upsilon_r\gamma\Delta_m + J_1\gamma^3\Delta_n + \varepsilon_\lambda \varpi_i 
\left(m\Delta_n+(n-1)\Delta_m\right)|f_m^{\rm c}|^{n-1}|f_n^{\rm c}|^m\cos\phi_f
\nonumber\\
&\mu = \varrho_r\gamma\Delta_n + J_2\gamma^3\Delta_m + \varepsilon_\mu \zeta_i 
\left(n\Delta_m+(m-1)\Delta_n\right)|f_m^{\rm c}|^n|f_n^{\rm c}|^{m-1}\cos\phi_f\nonumber\\
&\alpha = -\varepsilon_{\lambda}\tilde{\alpha}_i|f_m^{\rm c}|^{\frac{n-2}{2}}
|f_n^{\rm c}|^{\frac{m-1}{2}}\cos\Phi,\nonumber\\ 
&\beta = \,\varepsilon_{\lambda}\tilde{\beta}_i|f_m^{\rm c}|^{\frac{n-2}{2}}
|f_n^{\rm c}|^{\frac{m-1}{2}}\cos\Phi,\nonumber\\
&a=J_3\gamma,\nonumber\\
&b = J_4 \gamma -\frac{\tilde{\alpha}_i\tilde{\beta}_i}{2\varrho_r\gamma}
|f_m^{\rm c}|^{n-2}|f_n^{\rm c}|^{m-1}\sin^2\Phi ,\nonumber\\
&c = J_5 \gamma + \frac{\tilde{\alpha}_i^2}{2\upsilon_r\gamma}|f_m^{\rm c}|^{n-2}
|f_n^{\rm c}|^{m-1}\sin^2\Phi,\nonumber\\
&d = J_6 \gamma - \frac{\tilde{\alpha}_i\tilde{\beta}_i}{2\upsilon_r\gamma}
|f_m^{\rm c}|^{n-2}|f_n^{\rm c}|^{m-1}\sin^2\Phi.\nonumber\\
&e=J_7\gamma.
\label{eq:SWcoef}
\end{align}
Here $J_1, ..., J_7$ are ${\cal O}(1)$ functions of the TW coefficients~(\ref{eq:upsilon}-\ref{eq:coef2}) 
and are given in the Appendix, $\varepsilon_\lambda = {\rm sign}(\tilde{\lambda}_i)$, 
$\varepsilon_{\!\mu} = {\rm sign}(\tilde{\mu}_i)$, and 
\begin{align}
\Phi=\varepsilon_{\!\mu}\frac{\pi}{4}-\frac{\phi_f}{2}.  
\label{eq:Phi}
\end{align}
The coefficients~(\ref{eq:SWcoef}) are all real, expressed in terms of the unscaled 
parameters~(\ref{eq:upsilon}-\ref{eq:coef2}) and include the lowest order in 
$\epsilon$.  Additional interesting contributions, not necessarily at lowest order, have 
been kept as well.  In $\lambda$ and $\mu$, for example, we have included the 
leading order nonresonant (i.e., lacking explicit dependence on $m$, $n$, and $\phi_f$) 
coupling between the two forcings (the $J_1 \gamma^3\Delta_n$ term in $\lambda$ 
and the $J_2 \gamma^3\Delta_m$ term in $\mu$) and the lowest order ``resonant'' 
contribution  (the ${\cal O}(\epsilon^{m+n-1})$ terms varying as $\cos\phi_f$).  
In the nonlinear cross-coupling coefficients $b$, $c$, and $d$ there is an ${\cal O}(\epsilon^{m+n-4})$ 
resonant contribution coming from the damped SW modes that have been slaved away; this 
contribution appears at leading order when $m+n \leq 5$ but should be negligible for $m+n>5$. 

We now summarize the most important properties of the SW coefficients~(\ref{eq:SWcoef}).  All of 
this structure relies on the parameter symmetries ${\cal T}_s^{\rm par}$ and $\sigma^{\rm par}$, 
both possessed by the full hydrodynamical problem \cite{KumTuc94}, the restriction to weak damping 
($\gamma \sim |f_m^c| \sim |f_n^c| \ll 1$) and the assumed analytic parameter dependence 
on $\gamma$ and $f$.
\begin{itemize}

\item[(1)] For $m+n\geq 5$ all coefficients are ${\cal O}(\epsilon)$ or higher.  A similar 
result was found in the analytical expressions of Topaz and Silber \cite{TopSil02} and of 
Zhang and Vi\~nals \cite{ZhaVin97_2,Zha94}.  The latter authors point out \cite{ZhaVin97_1,Zha94} 
that the appearance of $\gamma$ in the nonlinear coefficients despite the fact that the 
Zhang-Vi\~nals equations contain only linear dissipation, should not be surprising.  
There is no reason why parameters in the linear terms of the governing equations cannot be involved 
in the nonlinear coefficients of the amplitude equations.   In our derivation this dependence is a 
natural consequence of the underlying temporal symmetries.   From a physical perspective it is 
also not surprising that the timescale for dissipative dynamics diverges as $1/\gamma$.

\item[(2)]  The unfolding parameters $\lambda$ and $\mu$ are linear combinations of 
$\Delta_m$ and $\Delta_n$, and are therefore rotated slightly (and stretched) with respect 
to the latter, ``more natural'' parameters.  This rotation is ${\cal O}(\epsilon^2)$ unless 
$m$:$n$ = 1:2 in which case it is ${\cal O}(\epsilon)$ and proportional to $\cos\phi_f$.  Similar 
rotation is discussed in more detail in \cite{TopSil02}, although those results are obtained 
away from the bicritical point.

\item[(3)] When $m$:$n$ = 1:2 the cubic cross-coupling coefficients $b$, $c$, and $d$ 
depend (sinusoidally) on the relative phase $\phi_f$ at leading order.  This behavior is 
similar to the leading order phase dependence found by Zhang and Vi\~nals~\cite{ZhaVin97_2}
for cubic cross-coupling coefficients with 1:2 forcing, a result obtained perturbatively and 
not from symmetry considerations.  Furthermore, the coefficients $b$, $c$, and $d$ diverge 
as $\epsilon \rightarrow 0$, indicating a failure of the reduction procedure in this limit.  This 
failure is not surprising because at $\gamma=0$ there are three additional critical modes 
which need to be included.   The three neglected SW modes satisfy $Z_j^+ \simeq -Z_j^-$ 
after the rescaling prior to Eqs.~(\ref{eq:SW1}-\ref{eq:SW3}) and experience ${\cal O}(\epsilon)$ 
damping.  The exceptionally strong coupling to these other modes when $m$:$n$ = 1:2 (the 
resonance terms of Eqs.~(\ref{eq:TWf1},\ref{eq:TWf3}), through which the coupling occurs, 
are {$\cal O$}($\epsilon$) or smaller unless $m$:$n$ = 1:2) helps explain the failure of the 
reduction procedure leading to Eqs.~(\ref{eq:SW1}-\ref{eq:SW3}).  In general, the description 
based on three SW modes is not expected to be valid if $\gamma$ is ``too'' small. 

\item[(4)] All measurable effects from coefficients~(\ref{eq:SWcoef}) are $2\pi$-periodic 
in $\phi_f$.  This is true even for $\alpha$ and $\beta$ (which appear to be $4\pi$-periodic 
in $\phi_f$; see Eq.~(\ref{eq:Phi})) because, owing to the invariance of 
Eqs.~(\ref{eq:SW1}-\ref{eq:SW3}) under $(A_j, \alpha,\beta) \rightarrow -(A_j, \alpha, \beta)$, 
only the {\it relative} sign of $\alpha$ and $\beta$ matters; the dynamics with $\Phi$ and 
$\Phi \pm \pi$ are equivalent.

\item[(5)] The resonant coefficients $\alpha$ and $\beta$ are of order $\epsilon^{(m+n-3)/2}$ 
and thus depend strongly on the choice of forcing frequencies, their influence diminishing 
exponentially with $m+n$; this fact helps explain why this type of resonant triad is not 
seen for large values of $m+n$ \cite{ArbFin02,ArbFin00}. The $\epsilon^{(m+n-3)/2}$ dependence is 
not universal however.  Near particular choices of $\phi_m$ and $\phi_n$ where $\cos\Phi=0$ 
the terms given in Eqs.~(\ref{eq:SWcoef}) vanish and, assuming the next nonvanishing term 
(not calculated) is a factor of $\gamma$ smaller, there is a cross-over to the more extreme 
$\epsilon^{(n+m-1)/2}$ dependence.  Conversely, one could say that resonance effects are 
{\it strongest} when $\Phi=0,\pi$.
\end{itemize}

%%%%%%%%%%%%%%%%%%%%%%%%%%%%%%%%%%%%
\subsubsection{Consequences of Hamiltonian structure}
\label{sssec:Ham}

In this section we examine the consequences of full Hamiltonian structure (not just the time reversal 
symmetry $\sigma^{\rm par}$) in the undamped (i.e., inviscid) problem; see 
\cite{ZhaVin97_1,Mil91,Zak68,Mile77,Bro74,LynAls97,Rad92}.  Specifically, we suppose  
that Eqs.~(\ref{eq:TWf1},\ref{eq:TWf3}) possess a Hamiltonian ${\cal H}$ with the 
complex amplitudes $Z_j^\pm$ obeying the equations of motion (see, e.g., \cite{Mile84,Kra94})
\begin{align}
\dot{Z_j^\pm} = \mp i \frac{\partial {\cal H}}{\partial \bar{Z}_j^\pm}, \qquad 
j = 1,2,3.\label{eq:defHam}
\end{align}
Requiring that ${\cal H}$ be a real scalar function, invariant under the symmetries 
${\cal T}_{\bs \phi}$, ${\cal R}$, $\kappa$, ${\cal T}_s^{\rm par}$, and $\sigma^{\rm par}$ leads to 
\begin{align}
{\cal H} =\,&h_1(|Z_1^\pm|^2 + |Z_2^\pm|^2) + h_2 |Z_3^\pm|^2 + 
h_3\big(\bar{f}_m (Z_1^+\bar{Z}_1^- + Z_2^+ \bar{Z}_2^-) +c.c.\big)\nonumber\\
+\,&h_4(\bar{f}_n Z_3^+ \bar{Z}_3^- +c.c.) + h_5\big(f_m^{n-1}\bar{f}_n^m (Z_1^+ \bar{Z}_1^- +
Z_2^+ \bar{Z}_2^-) + c.c.\big) + h_6(\bar{f}_m^n f_n^{m-1}Z_3^+ \bar{Z}_3^- + c.c.)\nonumber\\
+\,&h_7(\bar{f}_m^{\frac{n-2}{2}} f_n^{\frac{m-1}{2}} \bar{Z}_1^+ \bar{Z}_2^+ Z_3^+ + 
f_m^{\frac{n-2}{2}} \bar{f}_n^{\frac{m-1}{2}} \bar{Z}_1^- \bar{Z}_2^- Z_3^- +c.c.) + 
h_8(|Z_1^+|^2|Z_1^-|^2+|Z_2^+|^2|Z_2^-|^2)\nonumber\\
+\,&h_9(|Z_1^\pm|^4+|Z_2^\pm|^4) + h_{10}|Z_1^\pm|^2|Z_2^\pm|^2 + 
h_{11}|Z_1^\pm|^2|Z_2^\mp|^2 + h_{12} |Z_3^\pm|^2(|Z_1^\pm|^2+|Z_2^\pm|^2) \nonumber\\
+\,&h_{13} Z_1^\pm \bar{Z}_1^\mp \bar{Z}_2^\pm Z_2^\mp + 
h_{14}|Z_3^\pm|^4 + h_{15} |Z_3^\pm|^2(|Z_1^\mp|^2+|Z_2^\mp|^2) + 
h_{16}|Z_3^+|^2|Z_3^-|^2 \nonumber\\
+\,&h_{17}\big(f_m \bar{Z}_2^+ Z_2^-(|Z_1^+|^2+|Z_1^-|^2)+
f_m \bar{Z}_1^+ Z_1^-(|Z_2^+|^2+|Z_2^-|^2)+c.c.\big)\nonumber\\
+\,&h_{18}\big(f_m \bar{Z}_1^+Z_1^-(|Z_1^+|^2+|Z_1^-|^2)+f_m \bar{Z}_2^+ Z_2^-
(|Z_2^+|^2+|Z_2^-|^2)+c.c.\big)\nonumber\\
+\,&h_{19}\big(f_m(\bar{Z}_1^+ Z_1^- +\bar{Z}_2^+ Z_2^-)(|Z_3^+|^2+|Z_3^-|^2)+c.c.\big) + 
h_{20}\big(f_n Z_3^- \bar{Z}_3^+(|Z_3^+|^2+|Z_3^-|^2)+c.c.\big)\nonumber\\
+\,&h_{21}\big(f_n \bar{Z}_3^+ Z_3^-(|Z_1^+|^2+|Z_1^-|^2+|Z_2^+|^2+|Z_2^-|^2) +c.c.\big).
\label{eq:Ham}
\end{align}
Only those terms needed for comparison with Eqs.~(\ref{eq:TWf1},\ref{eq:TWf3}) have been 
included in Eq.~(\ref{eq:Ham}).  The coefficients $h_1, ..., h_{21}$ are real, and those terms involving a 
superscript $\pm$ should be understood as a sum over both signs, e.g., $|Z_1^\pm|^2 = |Z_1^+|^2 + 
|Z_1^-|^2$.  The equations of motion~(\ref{eq:defHam}) are equivalent to 
Eqs.~(\ref{eq:TWf1},\ref{eq:TWf3}) only if 
\begin{align}
&\tilde{\alpha}=\tilde{\beta},\quad {\rm e = g},\quad {\rm  h=f},\quad 
{\rm q_1=2q_2=2r_2},\quad {\rm q_3=q_4=r_1=r_3},\nonumber\\
&{\rm q_5=q_6=u_1=u_2},\quad {\rm r_4=r_5=s_1=s_2},\quad
{\rm s_3=2s_4=2u_3},
\label{eq:relHam}
\end{align} 
where all coefficients are evaluated at $\gamma=0$.  The most important of relations~(\ref{eq:relHam}) 
is $\tilde{\alpha}=\tilde{\beta}$ (i.e., $\tilde{\alpha}_i=\tilde{\beta}_i$) because that implies that 
the resonant coefficients $\alpha$ and $\beta$ of Eqs.~(\ref{eq:SWcoef}) have {\it opposite signs} 
for small $\gamma$.   Although this situation can change with increasing damping, the fact that there is 
good reason to expect $\alpha \beta <0$ has profound dynamical implications for 
Eqs.~(\ref{eq:SW1}-\ref{eq:SW3}) -- it is a prerequisite for a variety of interesting drifting solutions 
and heteroclinic cycles (see Section~\ref{sec:dyn}).   Chossat \cite{Cho01} has shown that the 
special form of advective nonlinearities leads to the same conclusion (resonant coefficients of opposite sign) 
for a general class of self-adjoint hydrodynamical problems (see also \cite{JonPro87}).   A strict interpretation 
of the equation $\tilde{\alpha}_i = \tilde{\beta}_i$ (which would also conclude that $\alpha$ and $\beta$ have 
equal {\it magnitudes}; see Eqs.~(\ref{eq:SWcoef})) is inappropriate here because, for one thing, such a  
statement depends on normalization conventions.    Provided $\alpha$ and $\beta$ do not vanish, the 
amplitudes $A_j$ may always be rescaled to set $|\alpha| = |\beta| = 1$.   More importantly, to say that  
the TW equations~(\ref{eq:TWf1},\ref{eq:TWf3}) display ``Hamiltonian structure" need not imply that 
the $Z_j^\pm$ and $\bar{Z}_j^\pm$ are {\it themselves} canonically conjugate Hamiltonian variables in 
the sense of  Eqs.~(\ref{eq:defHam}).    Transformations (simple rescalings, for example) that preserve 
dynamics should be allowed as well.   With this relaxed criteria we predict for the cubic coefficients simply  
that the oscillations of $b$, $c$, and $d$ with $\Phi$ are the same order in magnitude and that the oscillating 
part of $b$ and of $d$ (measured with respect to the nonresonant ${\cal O}(\epsilon)$ part) is strictly negative 
while that of $c$ is positive.  These oscillations appear at leading order only if $m$:$n$ = 1:2 or 3:2.  In the 
former case they are ${\cal O}(1/\epsilon)$ and can overwhelm the $\Phi$-independent 
${\cal O}(\epsilon)$ part.

%%%%%%%%%%%%%%%%%%%%%%%%%%%%%%%%%%%%
\subsubsection{Discussion of standing wave equations~(\ref{eq:SW1}-\ref{eq:SW3})}
\label{sssec:SWdiscussion}

We have related many properties of the coefficients of Eqs.~(\ref{eq:SW1}-\ref{eq:SW3}) 
to the influence of weakly broken temporal symmetries and Hamiltonian structure.  The spatial 
symmetries of the problem, however, remain fully intact and must be considered as well.  They act 
on the SW (cf.~Eqs.~(\ref{sym1:T}-\ref{sym1:R})) according to
\begin{align}
{\cal T}_{\bs \phi}&: A_j \rightarrow A_je^{i\phi_j}, \quad \bs{\phi}=(\phi_1,\phi_2), 
\quad \phi_1, \phi_2 \in [0,2\pi), \quad \phi_3=\phi_1+\phi_2,                                                                    
\label{sym2:T}\\
\kappa &: A_1 \leftrightarrow A_2,
\label{sym2:kappa}\\     
{\cal R}&: A_j \rightarrow \bar{A}_j,
\label{sym2:R}                                                                                                              
\end{align}
%]
In particular, the action of ${\cal R}$ explains why the coefficients in 
Eqs.~(\ref{eq:SW1}-\ref{eq:SW3}) are real.   Note that this nontrivial action of 
${\cal R}$ on the SW amplitudes, a necessary consequence of the spatial symmetries  
of the problem, allows for the possibility of a symmetry-breaking bifurcation to  
traveling (hereafter, drifting) waves.  Strictly speaking, the ``standing waves" of the 
previous section are standing only on the fast timescale set by $\omega$ (i.e., the 
eigenfunctions satisfy $|Z_j^+|=|Z_j^-|$).  On the slow timescale they may or may 
not drift, depending on whether or not they belong to the ${\cal R}$-invariant 
subspace (or a translation of it).

Recall that unless $n$ is even there are no quadratic resonant terms, i.e., $\alpha = \beta = 0$.  
This result, obtained in \cite{SilSke99}, follows directly from Eqs.~(\ref{eq:SW1}-\ref{eq:SW3}) 
by noting that the forced problem is invariant under a discrete time translation through one 
period of the forcing -- this is the extent to which time translation is usually considered.   
Accordingly, Eqs.~(\ref{eq:SW1}-\ref{eq:SW3}) must be equivariant under the operation 
$(A_1,A_2) \rightarrow (-1)^m(A_1,A_2)$, $A_3 \rightarrow (-1)^nA_3$.  If $n$ is even 
(and therefore $m$ odd) this discrete temporal symmetry is equivalent to the spatial translation 
${\cal T}_{(\pi,\pi)}$ and imposes no additional restrictions~\cite{SilSke99}.  However, if $n$ is 
odd the induced symmetry forces $\alpha = \beta = 0$.   Because we are interested in resonance 
effects we always take $n$ to be even, i.e., we assume $A_3$ is harmonic with respect to the 
overall forcing period $2\pi$. 

Even when $A_3$ is harmonic and the resonant coefficients are nonzero there is no guarantee that 
resonance effects can be easily observed.  As seen in Eqs.~(\ref{eq:SWcoef}), this depends strongly  
on $m+n$ and $\gamma$.  If $\alpha$ and $\beta$ are extremely small, their influence will be 
appreciable only while the system is extremely close to onset.  Although in one sense this limit is  
ideally-suited for the weakly nonlinear approach taken here, it may require far too 
much sensitivity for realistic experiments, i.e., the smallest available turn of the knob may 
immediately put the system into a regime dominated by cubic nonlinearities.  To estimate the 
likelihood of this occurring we characterize the range over which resonant terms 
contribute relative to the range over which the weakly nonlinear analysis is likely to be valid.  We may 
assume, for example, that a weakly nonlinear approach is ``justified" when $|A_j| \lesssim \epsilon_0$, 
where $\epsilon_0 = 0.1$ (or any other reasonable value).  The dynamics, however, will begin to 
switch from resonant (quadratic nonlinearities) to nonresonant (cubic nonlinearities) when these two 
effects are of the same order: $(\alpha,\beta)|A_j|^2 \sim (a,b,c,d,e) |A_j|^3$.  For 
concreteness, we suppose that this transition occurs when $|A_j| \sim \epsilon_{\rm res}
 \equiv {\rm max}(|\alpha|,|\beta|)/{\rm max}(|a|,|b|,|c|,|d|,|e|)$ and denote the ratio of 
$\epsilon_{\rm res}$ to $\epsilon_0$ by $\eta_{\rm res}$:
\begin{align}
\eta_{\rm res} = \frac{{\rm max}(|\alpha|,|\beta|)}{\epsilon_0\,{\rm max}(|a|,|b|,|c|,|d|,|e|)}.
\label{eq:etares}
\end{align}
Thus, resonance effects are expected to be important throughout the weakly nonlinear regime if 
$\eta_{\rm res} > 1$, while an earlier crossover to nonresonant behavior is anticipated 
when $\eta_{\rm res} < 1$.  Note that unless $m$:$n = 1$:2 (this case is more complicated 
because $b$, $c$, and $d$ diverge as $\epsilon \rightarrow 0$) $\eta_{\rm res}$ scales as 
$\epsilon^{(m+n-5)/2}$.  Broadly speaking, for small $\gamma$ (i.e., $\epsilon$) one expects 
that resonance effects will be easily observed when $m+n \leq 5$ but will become increasingly 
more delicate as $m+n$ increases.

%%%%%%%%%%%%%%%%%%%%%%%%%%%%%%%%%%%%
%%%%%%%%%%%%%%%%%%%%%%%%%%%%%%%%%%%%
\section{Numerical results using Zhang-Vi\~nals equations}
\label{sec:ZV}

In this section we test the predictions of Eqs.~(\ref{eq:SWcoef}) by numerically calculating the
SW coefficients from the Zhang-Vi\~nals model \cite{ZhaVin97_1,ZhaVin97_2}, a quasipotential 
formulation of the Faraday problem capturing the small-amplitude dynamics of deep fluid layers 
in the limit of weak damping.  This calculation also gives physically relevant coefficients which we 
use in Section~\ref{sec:dyn} to illustrate the SW dynamics.  We emphasize that the procedure 
described here is a direct reduction of the Zhang-Vi\~nals equations at the bicritical point to three SW  
amplitude equations, and does not rely on the symmetries of Section~\ref{ssec:unfold}.  
 
In the frame of the vibrated fluid we can write the effective gravitational acceleration in the form 
$g(t) = g_0 - g_m\cos(m\omega t + \phi_m) - g_n \cos(n\omega t + \phi_n)$, where $g_0>0$ 
is the usual gravitational acceleration.   Note that we are picking a convention by using  
cosines rather than sines, and by putting an overall minus sign in front of the applied 
acceleration.  Switching to the opposite sign convention is equivalent to taking 
$\phi_m \rightarrow \phi_m + \pi$, $\phi_n \rightarrow \phi_n + \pi$.  Because 
of Eq.~(\ref{eq:phif}), we then have $\phi_f \rightarrow \phi_f + (m-n)\pi$ and 
$\Phi \rightarrow \Phi -(m-n)\pi/2$.  Since most of the normal form coefficients in 
Eqs.~(\ref{eq:SWcoef}) depend on $\phi_f$ this convention does matter.  In $\alpha$, $\beta$, $b$, 
$c$, and $d$ it amounts to a shift by $\pi/2$ in the oscillations versus $\Phi$.  This follows from the fact 
that $m-n$ is odd, together with the invariance of the problem under $\Phi \rightarrow \Phi \pm \pi$ 
(i.e., $\phi_f \rightarrow \phi_f \mp 2\pi$; see Section~\ref{ssec:reductiontoSW}); changing the above 
sign convection is thus equivalent (for $\alpha$, $\beta$, $b$, $c$, and $d$) to taking 
$\Phi \rightarrow \Phi + \pi/2$.  The difference is observable -- in fact, many other patterns, 
e.g., hexagons, depend on a phase analogous to $\Phi$ (see \cite{PorSil02}) -- but, to our knowledge, 
even those experiments \cite{ArbFin02,EdwFau94,KudPieGol98,ArbFin00,Mul93} which 
investigate dependence on the temporal phase (equivalent to $\phi_f$) fail to report the convention 
used (one may infer the convention used in \cite{Mul93} from the agreement found in 
\cite{ZhaVin97_2}).   Without this information, such results are ambiguous.

The governing equations describing surface height $h({\bs x},\tau)$ and surface velocity potential 
$\varphi({\bs x},\tau)$ are expressed in nondimensional form as
\begin{align}
\partial_\tau h =\;&\gamma \nabla^2 h + \hat{\cal D}\varphi - 
\nabla\cdot(h \nabla \varphi)+\frac{1}{2}\nabla^2 (h^2 \hat{\cal D}\varphi) 
- \hat{\cal D}(h \hat{\cal D}\varphi)+\hat{\cal D}\big(h \hat{\cal D}
(h \hat{\cal D}\varphi)+\frac{1}{2}h^2 \nabla^2 \varphi \big),\nonumber \\
\partial_\tau \varphi =\; 
&\gamma\nabla^2\varphi + \Gamma_0 \nabla^2h-
G(\tau)h +\frac{1}{2}(\hat{\cal D}\varphi)^2 - \frac{1}{2}(\nabla \varphi)^2
\nonumber \\
-\,&(\hat{\cal D}\varphi)\big(h\nabla^2\varphi + \hat{\cal D}(h\hat{\cal D}
\varphi)\big) - \frac{1}{2}\Gamma_0\nabla\cdot\left((\nabla h)
(\nabla h)^2\right),
\label{eq:ZV}
\end{align}
where $G(\tau)=G_0-{\rm f}_m\cos(m\tau+\phi_m)-{\rm f}_n \cos(n\tau + \phi_n)$, 
and $({\rm f}_m,{\rm f}_n)= (g_m,g_n)k_0/\omega^2$; $\gamma$, $G_0$ and $\Gamma_0$ are as 
defined in Section~\ref{sec:sym}.   The nonlocal operator $\hat{\cal D}$ (see \cite{ZhaVin97_1}) multiplies 
each Fourier component by its wavenumber: $\hat{\cal D}\,{\rm exp}\{i{\bs k}\cdot {\bs x}\} = 
k \,{\rm exp}\{i {\bs k}\cdot {\bs x}\}$.

To extract coefficients from Eqs.~(\ref{eq:ZV}) we use a three-timing perturbation technique 
(see \cite{SilTopSke00} or \cite{SilSke99} for more details), introducing a bookkeeping parameter 
$\varepsilon$ and writing
\begin{align}
&h = \varepsilon\,h_1 + \varepsilon^2 h_2 + ...,\qquad \varphi = 
\varepsilon\,\varphi_1 + \varepsilon^2 \varphi_2 + ...,\\ 
&{\rm f}_m = {\rm f}_m^c + \varepsilon^2\,\tilde{\rm f}_m, \qquad
{\rm f}_n = {\rm f}_n^c +\varepsilon^2\,\tilde{\rm f}_n, \qquad
\partial_\tau \rightarrow \partial_\tau + \varepsilon \partial_{T_1} + 
\varepsilon^2 \partial_{T_2} + ...\,.
\end{align}
After $h_1$ and $\varphi_1$ are expressed in terms of Fourier modes, e.g., $h_1({\bf x},\tau) = \sum_{\bs k} 
p_k(\tau){\rm exp}\{i {\bs k}\cdot {\bs x}\}$ and $\varphi_1({\bf x},\tau) = 
\sum_{\bs k} q_k(\tau){\rm exp}\{i {\bs k}\cdot {\bs x}\}$, the ${\cal O}(\varepsilon)$ problem reduces to 
a series of uncoupled damped Mathieu equations parameterized by the wavenumber $k$:
\begin{align}
\partial_{\tau\tau}p_k+2\gamma k^2 \partial_\tau p_k + (\gamma^2 k^4 + G_0 k + \Gamma_0 k^3)p_k = 
k ({\rm f}_m\cos(m\tau + \phi_m)+{\rm f}_n \cos(n\tau + \phi_n)) p_k
\label{eq:mathieu}
\end{align} 
Note, from the first of Eqs.~(\ref{eq:ZV}), that $q_k = k^{-1} (\partial_\tau + \gamma k^2)p_k$.  At 
the bicritical point $({\rm f}_m,{\rm f}_n) = ({\rm f}_m^c,{\rm f}_n^c)$, there are exactly 
two periodic solutions to Eq.~(\ref{eq:mathieu}), $p_{k_1}(\tau)$ and $p_{k_3}(\tau)$, associated 
with the dual minima of the neutral stability curves (see Fig.~\ref{fig:23neut}).   

It is worthwhile at this point to extract the leading order linear TW coefficients in 
Eqs.~(\ref{eq:TWf1},\ref{eq:TWf3}) directly from Eq.~(\ref{eq:mathieu}).  First, by setting 
$\gamma=0$ and treating f$_m$ and f$_n$ as small perturbations, Eq.~(\ref{eq:mathieu}) can 
be solved with a multi-timing scheme that gives, as the first solvability condition,
\begin{align}
\dot{Z}_1^\pm = -i k_1/(2 m) {\rm f}_m^c Z_1^\mp,\qquad 
\dot{Z}_3^\pm = -i k_3/(2 n) {\rm f}_n^c Z_3^\mp.
\label{eq:fcoef}
\end{align}
Upon comparing Eqs.~(\ref{eq:fcoef}) with Eqs.~(\ref{eq:TWf1},\ref{eq:TWf3}) and 
(\ref{eq:coef2}) we identify
\begin{align}
\tilde{\lambda}_i = \frac{k_1}{2m}, \qquad \tilde{\mu}_i = \frac{k_3}{2n}.
\label{eq:lincoef1}
\end{align}
A complementary treatment with f$_m = {\rm f}_n =0$ and $\gamma$ in the role of small 
perturbation establishes, via Eqs.~(\ref{eq:TWf1},\ref{eq:TWf3}) and 
(\ref{eq:upsilon},\ref{eq:varrho}), that
\begin{align}
\upsilon_r = k_1^2, \qquad \varrho_r = k_3^2,
\label{eq:lincoef2}
\end{align}
Eqs.~(\ref{eq:lincoef1}) and (\ref{eq:lincoef2}) agree with the formulas given in \cite{TopSil02}.  
In Eqs.~(\ref{eq:lincoef2}) we see that the linear damping is proportional to $k^2$, as expected.   
The significance of Eqs.~(\ref{eq:lincoef1}), on the other hand, is in relation to the forcing convention 
discussed above.  We have found that $\tilde{\lambda}_i >0$ and $\tilde{\mu}_i > 0$, i.e., 
that $\varepsilon_\lambda = \varepsilon_{\!\mu} = 1$ in Eqs.~(\ref{eq:SWcoef},\ref{eq:Phi}).   
Had the alternative convention been used where $g(t) = g_0 + g_m\cos(m\omega t + \phi_m) + 
g_n \cos(n\omega t + \phi_n)$, we would have $\varepsilon_\lambda = \varepsilon_{\!\mu} = -1$.   

Since we are assuming $n$ is even and $m$ is odd, $p_{k_1}$ is subharmonic 
with respect to the forcing period $2\pi$: $p_{k_1}(\tau +2\pi) = - p_{k_1}(\tau)$, while 
$p_{k_3}$ is harmonic: $p_{k_3}(\tau +2\pi) = p_{k_3}(\tau)$.   To focus on resonant triads, 
we expand $h_1$ and $\varphi_1$ in terms of the three spatially 
resonant critical modes (see Fig.~\ref{fig:critcirc}):
\begin{align}
h_1 &= p_{k_1}(\tau)\big[A_1(T_1,T_2)e^{i {\bs k}_1\cdot {\bs x}} + 
A_2(T_1,T_2)e^{i {\bs k}_2\cdot {\bs x}}\big] + 
p_{k_3}(\tau)A_3(T_1,T_2)e^{i {\bs k}_3\cdot {\bs x}} + c.c.\\
\varphi_1 &= q_{k_1}(\tau)\big[A_1(T_1,T_2)e^{i {\bs k}_1\cdot {\bs x}} + 
A_2(T_1,T_2)e^{i {\bs k}_2\cdot {\bs x}}\big] + 
q_{k_3}(\tau)A_3(T_1,T_2)e^{i {\bs k}_3\cdot {\bs x}} + c.c.,
\end{align} 

At ${\cal O}(\varepsilon^2)$ and ${\cal O}(\varepsilon^3)$ one obtains differential 
equations from the solvability conditions.  These can be reconstituted by dropping 
$\varepsilon$ and setting $\partial_t = \partial_{T_1}+\partial_{T_2}$.  The subsequent 
equations are identical in form to Eqs.~(\ref{eq:SW1}-\ref{eq:SW3}), as they 
must be from symmetries~(\ref{sym2:T}-\ref{sym2:kappa}).  The explicit expressions for 
these coefficients are in general quite lengthy and must be computed numerically since 
they involve Fourier representations of $p_k(\tau)$ and $q_k(\tau)$.  For this reason we 
present only numerical results, concentrating mainly on the case $m$:$n$ = 3:2 for which 
there are many recent experimental results \cite{ArbFin02,ArbFin98}. 

\begin{figure}[ht]
\centerline{\includegraphics[width=4.5in]{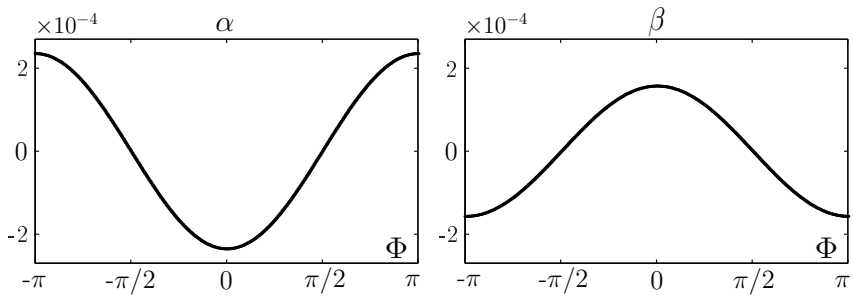}}
\caption{Dependence of $\alpha$ and $\beta$ on $\Phi$ in the case $m$:$n$ = 3:2 with 
physical parameters (in cgs units): $\rho=1$, $\nu=0.1$, $\Gamma=21.5$, and $\omega=10$, 
corresponding to $\gamma \simeq 2.08\!\times\!10^{-4}$.  To change $\Phi$ we set $\phi_n = 0$ 
and varied $\phi_m$.}
\label{fig:23resofPhi}
\end{figure}

\begin{figure}[ht]
\centerline{\includegraphics[width=5.3in]{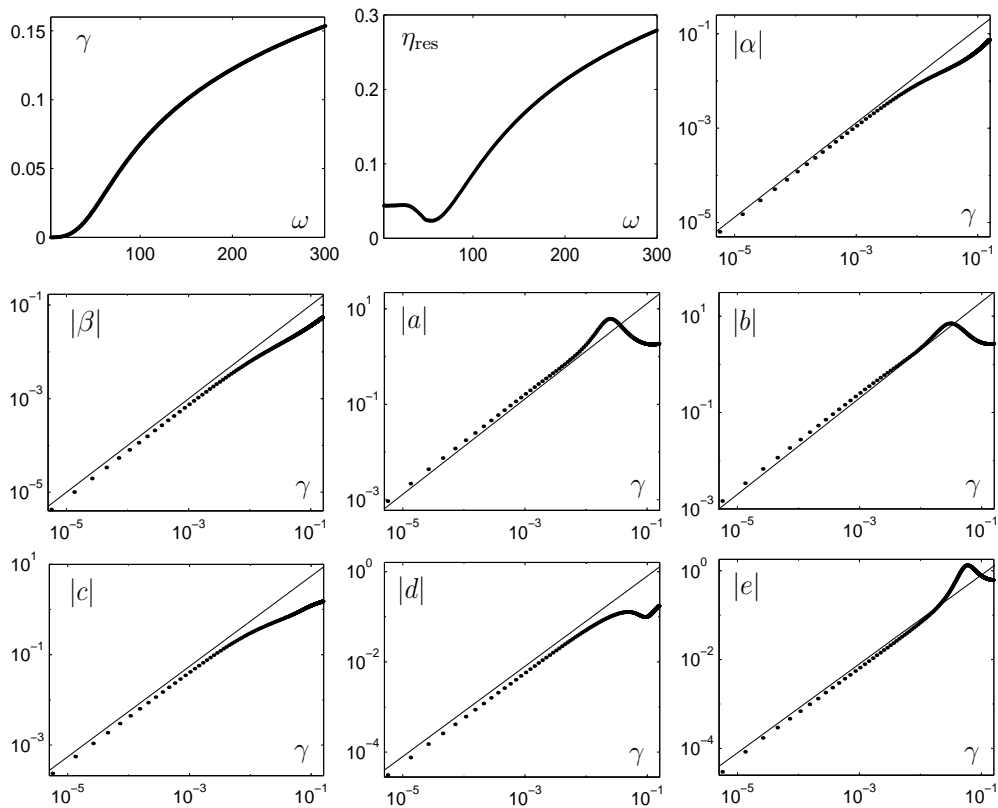}}
\caption{Behavior of $\gamma(\omega)$, $\eta_{\rm res}(\omega)$ (see Eq.~(\ref{eq:etares})),  
and the magnitude of the coefficients $|\alpha(\gamma)|$, $\dots$, $|e(\gamma)|$ of 
Eqs.~(\ref{eq:SW1}-\ref{eq:SW3}) over the interval $\omega \in [3,300]$ rad/s with 
$\phi_m = -\pi/4$, $\phi_n=0$, $\rho=1$, $\nu=0.1$, and $\Gamma=21.5$ (cgs units).  
The straight lines show the predicted (linear) $\gamma$ dependence.  In the calculation 
of $\eta_{\rm res}$ we use $\epsilon_0 = 0.1$.}
\label{fig:23unscalemax}
\end{figure}

When $m$:$n$ = 3:2 the coefficients $\alpha$, $\beta$, $b$, $c$, and $d$ depend at leading 
order on $\Phi$.   We find, however, that for the Zhang-Vi\~nals equations the $\Phi$-independent 
parts of $b$, $c$, and $d$ are considerably larger than the $\Phi$-dependent parts despite the fact 
that they formally appear at the same order in $\gamma$.  The most important effect of  
$\Phi$ is therefore in the $\cos \Phi$ behavior of $\alpha$ and $\beta$ (see Eqs.~(\ref{eq:SWcoef})) 
which we confirm in Fig.~\ref{fig:23resofPhi}.      
%%%%%%%%%%%%%%%%%%%%%%%%%%%%%%%%%%%%%%%%FIG
%%%%%%%%%%%%%%%%%%%%%%%%%%%%%%%%%%%%%%%%FIG
Observe that $\alpha(\Phi)$ and $\beta(\Phi)$ have the opposite phase, i.e., $\alpha \beta < 0$, 
as predicted by the arguments of Section~\ref{sssec:Ham}.  Maximal values of $|\alpha|$ 
and $|\beta|$ are realized (for small $\gamma$) when $\Phi =0$.  We use such an ``optimal'' 
choice of phases, e.g., $\phi_m=-\pi/4$ and $\phi_n = 0$, to test in Fig.~\ref{fig:23unscalemax} 
%%%%%%%%%%%%%%%%%%%%%%%%%%%%%%%%%%%%%%%%FIG
%%%%%%%%%%%%%%%%%%%%%%%%%%%%%%%%%%%%%%%%FIG
the scalings predicted by Eqs.~(\ref{eq:SWcoef}).  Note that, as expected for $m$:$n$ = 3:2, 
we find $\eta_{\rm res} \sim 0.02$--0.3, and resonance effects are likely to be important over 
much of the weakly nonlinear regime.  

\begin{figure}[ht]
\centerline{\includegraphics[width=5.3in]{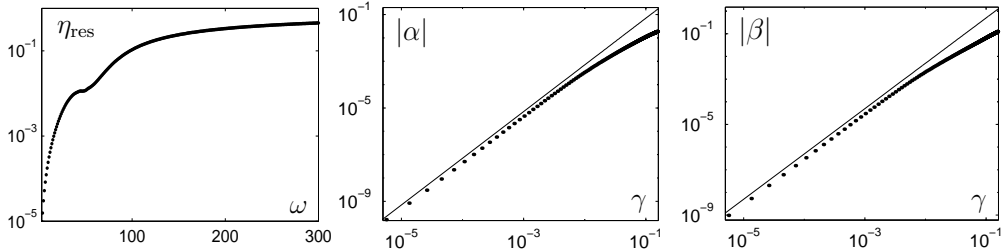}}
\caption{Dependence of $\eta_{\rm res}(\omega)$ (using $\epsilon_0 = 0.1$), 
$|\alpha(\gamma)|$, and $|\beta(\gamma)|$ when $\Phi = \pi/2$.  Parameters are 
the same as in Fig.~\ref{fig:23unscalemax} except with $\phi_m=\pi/4$.  Straight 
lines show the predicted $\gamma^2$ dependence.}
\label{fig:23unscalemin}
\end{figure}

For values of $\Phi$ near $\pm \pi/2$, the $\gamma$ dependence of $\alpha$ and $\beta$ switches 
from the exponent $(m+n-3)/2$, characteristic of the terms given in Eqs.~(\ref{eq:SWcoef}), to 
the exponent $(n+m-1)/2$, reflecting the next nonvanishing term (not calculated).  
Fig.~\ref{fig:23unscalemin} illustrates this second type of behavior by showing
%%%%%%%%%%%%%%%%%%%%%%%%%%%%%%%%%%%%%%%%FIG
%%%%%%%%%%%%%%%%%%%%%%%%%%%%%%%%%%%%%%%%FIG
$\eta_{\rm res}(\omega)$, $|\alpha(\gamma)|$, and $|\beta(\gamma)|$ when $\Phi=\pi/2$.  
The remaining coefficients in Eqs.~(\ref{eq:SW1}-\ref{eq:SW3}) are essentially unchanged 
from Fig.~\ref{fig:23unscalemax}.  Note that $\eta_{\rm res} \rightarrow 0$ as $\omega \rightarrow 0$ 
(i.e., $\gamma \rightarrow 0$), indicating that resonance effects are becoming increasingly weak -- 
larger values of $\omega$ are required to see them more easily.  Furthermore, besides the fact that 
$\alpha$ and $\beta$ are smaller, we find that $\alpha \beta >0$ over the calculated range 
$\omega \in [3,300]$.  This is in sharp contrast to the typical case, with $\Phi$ bounded away from 
$\pm \pi/2$, when the leading order terms given in Eqs.~(\ref{eq:SWcoef}) dominate and 
$\alpha \beta <0$.   Because the relative phase $\phi_f$ (equivalently, $\Phi$) can alter the sizes 
of $\alpha$ and $\beta$ and the sign of their product, it can have a dramatic effect on the dynamics 
of resonant triads, as shown in the following section.

%%%%%%%%%%%%%%%%%%%%%%%%%%%%%%%%%%%%%%%%%%%
%%%%%%%%%%%%%%%%%%%%%%%%%%%%%%%%%%%%%%%%%%%
\section{Dynamics}
\label{sec:dyn}
In this section we discuss the dynamics of Eqs.~(\ref{eq:SW1}-\ref{eq:SW3}) with an eye to 
the influence of relations~(\ref{eq:SWcoef}) and (\ref{eq:relHam}).    As shown by Guckenheimer 
and Maholov~\cite{GucMah92} these dynamics can be very rich, in part because a number of solutions 
drift or ``travel" on the slow time scale, as discussed in Section~\ref{sssec:SWdiscussion}.    In the 
following we refer to these as drifting waves (DW) to emphasize that this motion is much slower than 
the fast (standing) oscillations associated with the parametric forcing.

While the temporal symmetries are now implicit, expressed in relations~(\ref{eq:SWcoef}) and 
(\ref{eq:relHam}) describing the scaling, parameter-dependence, and relative sign of the SW 
normal form coefficients, the unbroken spatial symmetries~(\ref{sym2:T}-\ref{sym2:kappa}) 
force the existence of dynamically invariant subspaces containing distinct classes of solutions; 
the most important of these {\it fixed point subspaces} are listed in Table~\ref{tb:fixptsub}.
\begin{table}[ht] \begin{center}
\begin{tabular}{|c|c|c|}
\hline 
Subspace & \hspace{.05in} Fixed by  \hspace{.05in} & Defining property \\
\hline
${\cal U_R}$ & $\cal{R}$ & ${\rm Im}(A_j)=0$ \\
\hline
${\cal U}_\kappa$ & $\kappa$  & $A_1=A_2$  \\
\hline
${\cal U}_1$ & ${\cal T}_{(0,\phi)}$  & $A_2=A_3=0$ \\
\hline
${\cal U}_2$ & ${\cal T}_{(\phi,0)}$  & $A_1=A_3=0$ \\
\hline
${\cal U}_3$ & $\kappa$, ${\cal T}_{(\phi,-\phi)}$ & $A_1=A_2=0$ \\
\hline
\end{tabular} 
\caption{Invariant subspaces of Eqs.~(\ref{eq:SW1}-\ref{eq:SW3}).}
\label{tb:fixptsub} 
\end{center}
\end{table}
Not all of these subspaces are isolated and independent.  For example, $\cal{U_R}$ contains the 
solutions that are invariant under inversion through the origin (i.e., under ${\cal R}$) but there is 
a {\it torus} of equivalent subspaces generated by ${\cal T}_{\bs \phi}$; each of these 
related subspaces is invariant under inversion through an appropriate (translated) origin.  
Similarly, there is a {\it circle} of subspaces equivalent to ${\cal U}_\kappa$ generated by 
${\cal T}_{(\phi,-\phi)}$; solutions in these subspaces satisfy $A_2=e^{i2\phi}A_1$ and are 
reflection-symmetric about a line parallel to $\bs{k_3}$.   The subspaces ${\cal U}_1$ and 
${\cal U}_2$ are related by the symmetry $\kappa$ and are equivalent.   In the following 
we use ${\cal U_R}$, ${\cal U}_\kappa$, and ${\cal U}_1$ to denote the entire family of 
equivalent subspaces, the representative members of which are listed in Table~\ref{tb:fixptsub}.   
In practice, the most important subspaces are ${\cal U_R}$ and ${\cal U}_\kappa$ because 
they relate to easily recognizable characteristics:  solutions in ${\cal U_R}$ are strictly 
standing (i.e., they do not drift; see below) while solutions in ${\cal U}_\kappa$ possess a 
line of symmetry and may drift only in a direction parallel (or antiparallel) to $\bs{k_3}$.

For most purposes it helps to factor out the continuous translation symmetry 
${\cal T}_{\bs \phi}$ by writing $A_j=a_je^{i\theta_j}$ and introducing 
the ${\cal T}_{\bs \phi}$-invariant phase $\theta=\theta_3-\theta_1-\theta_2$.  One 
then obtains the four-dimensional system
\begin{align}
\dot{r}_1 &= \lambda r_1 + \alpha r_2 r_3\cos\theta + 
r_1(ar_1^2+br_2^2+cr_3^2),\label{eq:polar1}\\
\dot{r}_2 &= \lambda r_2 +\alpha  r_1 r_3\cos\theta + 
r_2(ar_2^2+br_1^2+cr_3^2),\label{eq:polar2}\\
\dot{r}_3 &= \mu r_3 + \beta r_1 r_2\cos\theta + 
r_3(dr_1^2+dr_2^2+er_3^2),\label{eq:polar3}\\
\dot{\theta} &= -\big(\beta\, \frac{r_1 r_2}{r_3} + \alpha \frac{r_2r_3}{r_1} + 
\alpha \frac{r_1r_3}{r_2}\big)\sin\theta,\label{eq:polar4}
\end{align}
where the $\theta_j$ satisfy
\begin{align}
\dot{\theta_1} = \alpha \frac{r_2r_3}{r_1} \sin\theta, \qquad
\dot{\theta_2} = \alpha \frac{r_1r_3}{r_2} \sin\theta, \qquad
\dot{\theta_3} = -\beta \frac{r_1r_2}{r_3}\sin\theta .\label{eq:thetaj}
\end{align}
Eqs.~(\ref{eq:thetaj}) guarantee that solutions in ${\cal U_ R}$, which corresponds to 
$\{\theta=0\}$ $\cup$ $\{\theta=\pi\}$ in Eqs.~(\ref{eq:polar1}-\ref{eq:polar4}), are also 
{\it standing} (i.e., all phase velocities are zero).    Furthermore, from Eq.~(\ref{eq:polar4}) 
it follows that unless $\alpha\beta <0$ the quantity multiplying $\sin\theta$ 
is single-signed and all trajectories approach ${\cal U_ R}$ (see \cite{GucMah92,ArmGucHol88}), i.e.,  
there are no permanently drifting solutions if $\alpha\beta >0$.  The fact that one can {\it expect} 
drifting solutions on the basis of underlying Hamiltonian structure was the conclusion of 
Section~\ref{sssec:Ham}
  
%%%%%%%%%%%%%%%%%%%%%%%%%%%%%%%%%%%%%%%%%%%%%%%
\subsection{Principal solutions}               
We describe here the most important solutions of Eqs.~(\ref{eq:SW1}-\ref{eq:SW3}), 
equivalently, Eqs.~(\ref{eq:polar1}-\ref{eq:polar3}).   Many of these solutions (or their 
analogs) are discussed in \cite{GucMah92}.  They are summarized, along with the bifurcations 
they undergo, in Table~\ref{tb:bifs}.          
\begin{table}[ht]
\centerline{\includegraphics{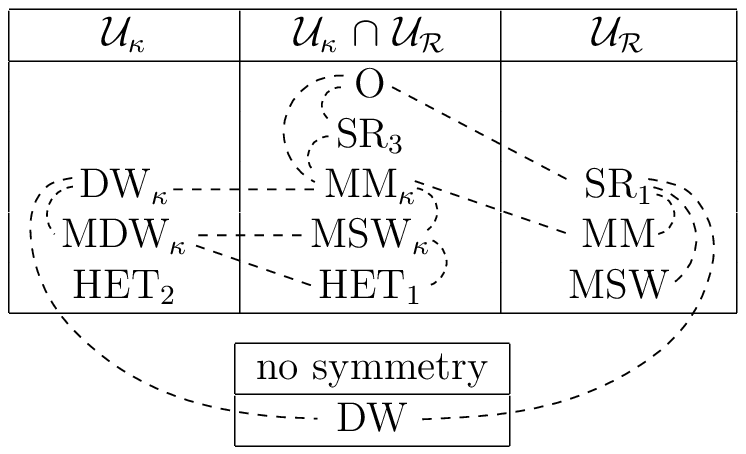}}
\caption{Typical solutions (see text) classified according to the reflection symmetries 
${\cal R}$ and $\kappa$.  Broken lines indicate potential bifurcations.}
\label{tb:bifs}
\end{table}

\noindent
(1)  Standing rolls SR$_1$ $\in$ ${\cal U}_1$ 

\noindent
The SR$_1$ states satisfy $|A_1|^2=-\lambda/a$, $A_2=A_3=0$.   Physically, they represent 
SW with dominant frequency $m/2$ described by a single wavevector ${\bs k}_1$.   
Standing rolls SR$_2$ with wavevector ${\bf k}_2$ are related to SR$_1$ through  
the symmetry $\kappa$ and are equivalent.   Both bifurcate from the flat 
state (O) when $\lambda=0$, supercritically if $a<0$.   The eigenvalues of SR$_1$ within the 
two-dimensional subspace ${\cal U}_1$ are 0 (a result of the translational symmetry 
${\cal T}_{\bs \phi}$) and $-2\lambda$.  Transverse to ${\cal U}_1$ the eigenvalues $\xi$ have 
multiplicity two and satisfy
\begin{align}
a^2\xi^2 - a \left(a \mu+ \lambda(a-b-d)\right) \xi + \lambda(a-b)(a\mu-d\lambda)+ a\beta\lambda = 0.
\end{align} 
Near onset (i.e., $|\lambda|\ll$ 1) these eigenvalues are approximated by $\mu$ and 
$\lambda(a-b+\alpha\beta/\mu)/a$.  Thus, the standing rolls SR$_1$ are stable at onset if 
$a<0$ (the bifurcation is supercritical), $\mu<0$ (stability with respect to perturbations of 
type $A_3$ is inherited from the basic flat state), and $\mu (b-a)>\alpha\beta$; they are 
unstable otherwise.   

The SR$_1$ states undergo a steady-state bifurcation along the line
\begin{align}
\mu = \frac{\alpha\beta}{b-a} + \frac{d}{a}\, \lambda,
\label{eq:SR1ssbif}
\end{align}
giving rise to a branch of mixed modes (MM) contained in ${\cal U_R}$ with $A_2 \neq 0$, 
$A_3 \neq 0$, and experience a Hopf bifurcation along the line 
\begin{align}
\mu = \frac{b+d-a}{a}\, \lambda,
\label{eq:SR1Hopf}
\end{align}  
provided $\alpha\beta < \lambda(a-b)^2/a<0$.    This Hopf bifurcation with $O(2)$ symmetry 
(see, e.g., \cite{GolSteSha88})) {\it simultaneously} produces modulated standing waves (MSW) 
contained in ${\cal U_R}$, and DW not in ${\cal U_R}$.    Note that $\alpha\beta <0$ is a necessary 
condition for this bifurcation, as it must be if DW are produced.   Assuming this condition holds, the line 
of Hopf bifurcations~(\ref{eq:SR1Hopf}) intersects the line of steady bifurcations~(\ref{eq:SR1ssbif}) 
at the Takens-Bogdanov point:  $(\lambda,\mu)=\alpha\beta(a,b+d-a)/(b-a)^2$.

\noindent
(2)  Standing Rolls SR$_3$ $\in$ ${\cal U}_3$.

\noindent
The SR$_3$ states satisfy $A_1=A_2=0$, $|A_3|^2=-\mu/e$ and represent harmonic SW 
(dominant frequency $n/2$) with wavevector ${\bs k}_3$.  They bifurcate from O 
when $\mu=0$, supercritically if $e<0$.   Within ${\cal U}_3$ the eigenvalues 
are 0 and $-2\mu$ while the remaining four eigenvalues come in pairs given by
\begin{align}
s = \lambda - \frac{c}{e}\mu \pm \alpha \sqrt{\frac{-\mu}{e}}.
\label{eq:SR3eig}
\end{align}
In contrast to SR$_1$, only steady-state bifurcations can occur on SR$_3$ and these generate 
branches of mixed modes MM$_\kappa$ $\in$ ${\cal U}_\kappa$ (i.e., the unstable eigenvectors 
satisfy $|A_1|=|A_2|$).\\

\noindent
(3)  $\kappa$-symmetric mixed modes MM$_\kappa$ $\in$ ${\cal U_R}$ $\cap$ ${\cal U}_\kappa$ .

\noindent
These mixed modes bifurcate, along with SR$_1$, from O when $\lambda=0$.  The 
${\cal T}_{\bs \phi}$-invariant phase $\theta$ must be 0 or $\pi$ while the amplitudes satisfy
\begin{align}
&(2cd-(a+b)e)r_3^3 \pm (2\alpha d+\beta c)r_3^2 + (2d\lambda+\alpha \beta-(a+b)\mu)r_3 
\pm \beta \lambda = 0, \nonumber\\
&r_1^2=r_2^2=-\frac{cr_3^2\pm \alpha r_3+\lambda}{a+b},
\label{eq:MMkappa}
\end{align}
with $\cos\theta$ dictating the $\pm$ sign.  Eqs.~(\ref{eq:MMkappa}) show that there can be up to 
three distinct MM$_\kappa$ states for each of the two possible $\theta$ values.  The additional solutions 
arise in saddle-node bifurcations, and other transitions can occur as well.   Hopf bifurcations 
lead to modulated standing waves, MSW$_\kappa$, still contained in ${\cal U}_\kappa$ $\cap$ 
${\cal U_R}$, while symmetry-breaking bifurcations lead to drifting waves DW$_\kappa$ $\in$
${\cal U}_\kappa$ (broken ${\cal R}$ symmetry) or to mixed modes MM $\in$ ${\cal U_R}$ (broken 
$\kappa$ symmetry).  The MM states bifurcate only from the MM$_\kappa$ branch with 
$\cos\theta = {\rm sign}(\alpha (a-b))$ and do so when
\begin{align}
&\mu = \frac{\alpha\beta}{b-a} + \frac{e}{c}\lambda + \frac{2 \alpha (cd-ae)}{c^2(a-b)^2}
\big(a\alpha \pm \sqrt{a^2\alpha^2-\lambda c(b-a)^2}\,\big).
\end{align}

\noindent
(4)  $\kappa$-symmetric drifting waves DW$_\kappa$ $\in$ ${\cal U}_\kappa$.

\noindent
These solutions require $\alpha\beta <0$.  Within Eqs.~(\ref{eq:SW1}-\ref{eq:SW3}) they drift 
with constant velocity parallel to $\pm{\bs k}_3$ (i.e., $\dot{\theta}_1 = \dot{\theta}_2 = \dot{\theta}_3/2$).    
Within the four-dimensional reduced system~(\ref{eq:polar1}-\ref{eq:polar4}) they appear as 
fixed points with $\theta \neq 0$ or $\pi$ satisfying
\begin{align}
&r_3^2=-\frac{\beta(\mu+2\lambda)}{4\alpha(a+b+d)-\beta(2c+e)}, \qquad 
r_1^2=r_2^2=-\frac{2\alpha}{\beta}\,r_3^2, \nonumber\\
&r_3 \cos\theta = \frac{\mu(2\alpha(a+b)-\beta c)+\lambda(\beta e -4\alpha d)}
{\alpha(4\alpha(a+b+d)-\beta(2c+e))},
\end{align}
and exist in the region
\begin{align}
(\mu(2\alpha(a+b)-c\beta)-\lambda (4d\alpha-e\beta))^2 \leq 
(\mu+2\lambda)\alpha^2(4\alpha\beta(a+b+d)-\beta^2(2c+e)).
\end{align}
The DW$_\kappa$ states themselves can undergo Hopf bifurcations to create modulated drifting 
waves MDW$_\kappa$ $\in$ ${\cal U}_\kappa$, as well as symmetry-breaking bifurcations to 
DW not in ${\cal U}_\kappa$.   This latter bifurcation occurs along the line
\begin{align}
\mu = \frac{4\alpha (b+d-a)-\beta e}{4\alpha a- \beta c}\, \lambda.
\end{align}

\noindent
(5)  Mixed modes MM $\in$ ${\cal U_R}$

\noindent
These mixed states (without $\kappa$ symmetry) satisfy 
\begin{align}
&r_3^2 = \frac{\alpha \beta a + (a\mu-d\lambda)(a-b)}{(a-b)(cd-ae)},\qquad 
\cos\theta = {\rm sign}(\alpha(a-b)),\nonumber\\
&r_1^2+r_2^2 = -\frac{\lambda+c r_3^2}{a}, \qquad r_1r_2 =\left|\frac{\alpha}{a-b}\right| r_3.
\end{align}
They form a branch of solutions connecting the SR$_1$ (SR$_2$) states with the MM$_\kappa$ 
branch having $\cos\theta = {\rm sign}(\alpha(a-b))$.  There are no saddle-node bifurcations 
but if $\alpha\beta < 0$ there is an ${\cal R}$ symmetry-breaking instability that creates drifting waves 
DW along the line
\begin{align}
\mu = \frac{e}{c}\,\lambda + \frac{\alpha \beta(bc+cd-ac-ae)}{c(a-b)^2}.
\end{align}

\noindent
(6) Drifting waves DW.

\noindent
These steadily drifting solutions (requiring $\alpha\beta < 0$) of 
Eqs.~(\ref{eq:SW1}-\ref{eq:SW3}) are not in ${\cal U}_\kappa$ and are therefore free to 
move in any direction (i.e., their drift direction is determined by the coefficients, not by symmetry).  
They are represented by fixed points within Eqs.~(\ref{eq:polar1}-\ref{eq:polar4}) satisfying 
\begin{align}
&r_3^3=\frac{a\mu +(a-b-d)\lambda}{bd+cd-ac-ae}, \qquad
r_1^2+r_2^2=-\frac{\lambda + cr_3^2}{a}, \nonumber\\
&\cos^2\theta = \frac{(a-b)^2(\lambda + cr_3^2)}{\alpha\beta a}, \qquad 
r_1r_2 = \frac{\alpha r_3\cos\theta}{a-b}.
\end{align}
As with MM (and DW$_\kappa$) there are no saddle-node bifurcations.   In fact, for the 
sets of parameters we investigated there were no additional bifurcations on the DW branch.\\

\noindent
(7) $\kappa$-symmetric modulated standing waves MSW$_\kappa$ $\in$ ${\cal U_R}$ 
$\cap$ ${\cal U}_\kappa$.

\noindent
These states appear via Hopf bifurcation from MM$_\kappa$.  The form of the bifurcation set 
is nontrivial but reduces to a parabola (see \cite{Dan86,ProJon88}) near the origin $(\mu,\lambda) = (0,0)$:
\begin{align}
\mu \simeq -\frac{3 e}{|\alpha|}\lambda^2, \qquad  \alpha\beta e \lambda < 0
\end{align}
The MSW$_\kappa$ solutions may undergo an ${\cal R}$ symmetry-breaking bifurcation 
to modulated drifting waves MDW$_\kappa$ and are typically destroyed in a global bifurcation  
involving a heteroclinic connection between O and SR$_3$ (see \cite{ArmGucHol88,PorKno01}).\\

\noindent
(8)  Modulated standing waves MSW $\in$ ${\cal U_R}$ .

\noindent
These solutions do not have the $\kappa$ symmetry of MSW$_\kappa$.  They appear, along with 
DW, in a Hopf bifurcation on SR$_1$.\\

\noindent
(9) $\kappa$-symmetric modulated drifting waves MDW$_\kappa$ $\in$ ${\cal U}_\kappa$.

\noindent
These three-frequency states can emerge from DW$_\kappa$ through Hopf bifurcation 
or from MSW$_\kappa$ via an ${\cal R}$ symmetry-breaking bifurcation.\\

\noindent
(10) Heteroclinic cycles HET.

\noindent
There are (at least) three kinds of heteroclinic cycles of relevance to 
Eqs.~(\ref{eq:SW1}-\ref{eq:SW3}).  Structurally stable heteroclinic cycles (HET$_1$) 
connecting two SR$_3$ solutions, out of phase by $\pi$ (i.e., related by a translation 
${\cal T}_{(\phi, \pi-\phi)}$), have been well studied \cite{JonPro87,ArmGucHol88} in the context of the 
1:2 steady-state spatial resonance with O(2) symmetry and in resonant triads as 
well~\cite{GucMah92}.  They are generic solutions when $\alpha \beta <0$ that appear at small 
amplitude ($|\lambda|, |\mu| \ll 1$). 

\begin{figure}[ht]
\centerline{\includegraphics{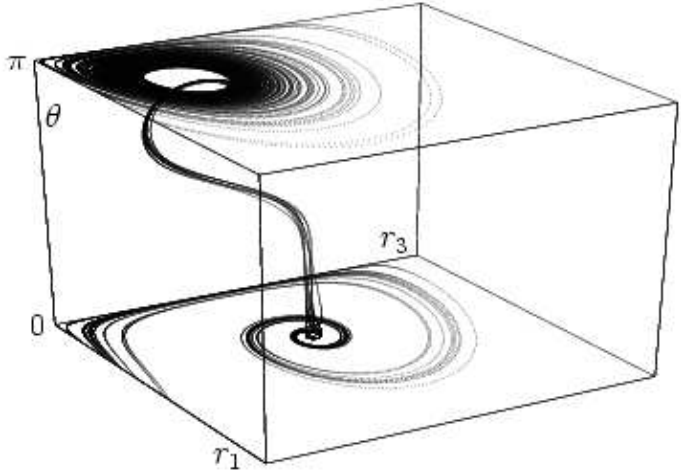}}
\caption{Chaotic attractor in Eqs.~(\ref{eq:polar1}-\ref{eq:polar4}) obtained at $(\lambda, 
\mu)=(-0.0377,0.035)$ with the coefficients $\alpha=-0.19$, $\beta=0.166$, $a=-0.0399$, $b=-0.436$, 
$c=1.27$, $d=-1.43$, and $e=-0.106$, derived from the Zhang-Vi\~nals model~(\ref{eq:ZV}) 
with $m$:$n = 1$:2,  $\Phi=\pi/4$, and (in cgs units) $\omega=200$, $\nu=0.01$, $\Gamma=70$.  
This attractor is in ${\cal U}_\kappa$ and visits the neighborhood of two MM$_\kappa$ fixed 
points: one with $\theta=0$ and one with $\theta=\pi$.  The latter is near a Hopf bifurcation 
to MSW$_\kappa$.  The boundary $r_3=0$ should be identified at $\theta=0$ and $\theta=\pi$.}
\label{fig:12chaotic}
\end{figure}

In addition, there are much more intricate heteroclinic cycles (HET$_2$) involving connections 
among O, SR$_3$, MSW$_\kappa$, and MM$_\kappa$ (see \cite{PorKno01}).  Under appropriate 
conditions, these cycles form near the HET$_1$ cycles (in a neighboring region of parameter space 
but further from the origin).  The associated dynamics is organized by a sequence of transitions 
among distinct heteroclinic cycles (one of which is structurally stable) and contains chaos of 
Shil'nikov type.    The type of chaotic attractor one typically finds is shown in Fig.~\ref{fig:12chaotic}.
%%%%%%%%%%%%%%%%%%%%%%%%%%%%%%%%%%%%%%%%FIG
%%%%%%%%%%%%%%%%%%%%%%%%%%%%%%%%%%%%%%%%FIG

When the resonance terms vanish (i.e., at $\alpha=\beta=0$) there are, under certain conditions 
\cite{GucMah92}, heteroclinic cycles connecting SR$_1$, SR$_2$, and SR$_3$.  Because of the resonant 
terms in Eqs.~(\ref{eq:SW1}-\ref{eq:SW3}) these cycles are no longer present.   Nonetheless, for large 
$m+n$ when resonance effects are diminished, or at values of $\phi_m$ and $\phi_n$ where 
$\Phi=\pm \pi/2$ and the leading order resonant term vanishes (see Eqs.~(\ref{eq:SWcoef})), 
these ``nearby" cycles may yet exert an influence on the dynamics. 

%%%%%%%%%%%%%%%%%%%%%%%%%%%%%%%%%%%%%%%%%%%%%
\subsection{Bifurcations: numerical examples}
\label{ssec:bifurcations}

\begin{figure}[ht]
\centerline{\includegraphics[width=4in]{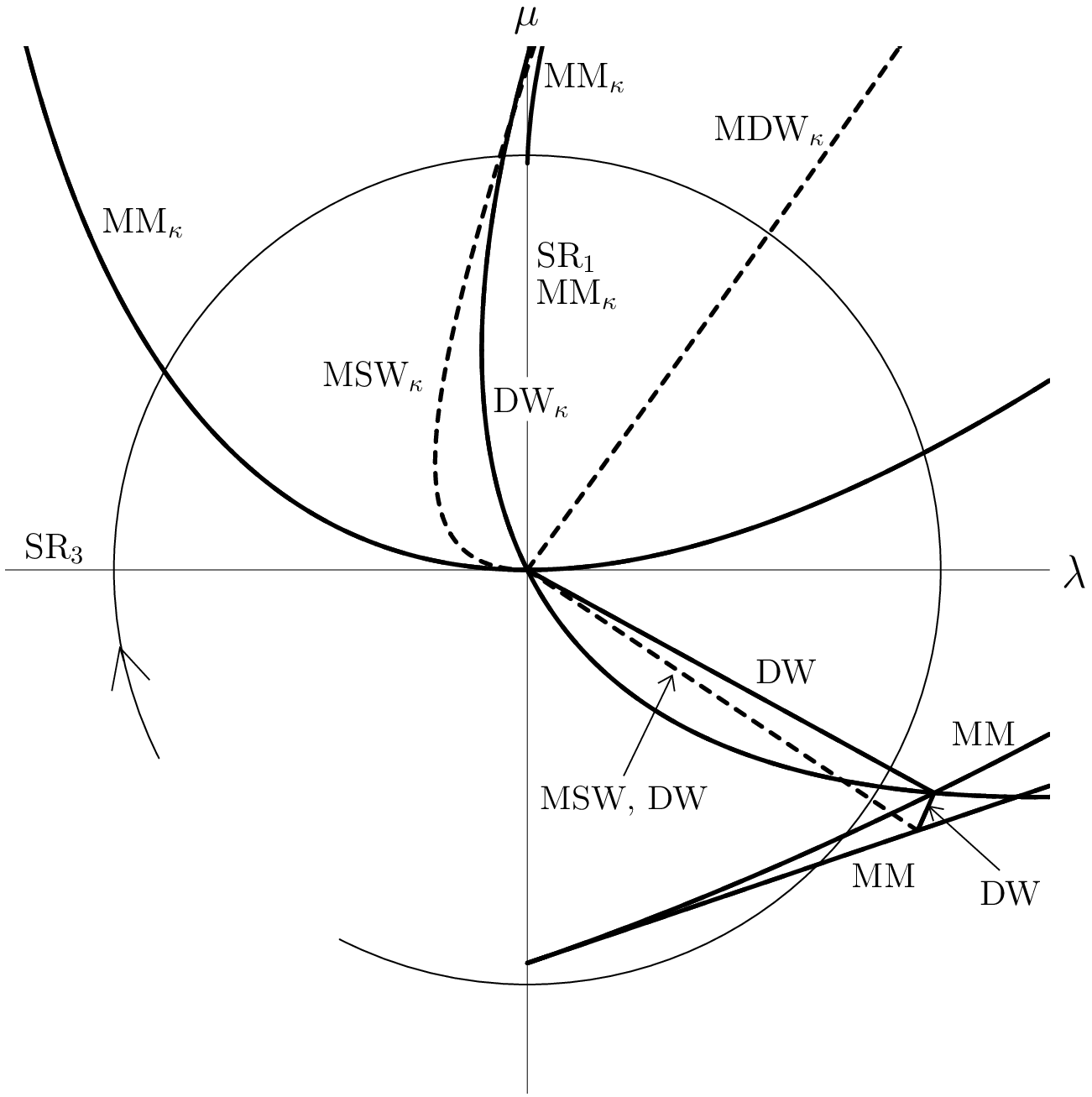}}
\caption{Bifurcation sets for the case $m$:$n = 1$:2 and $\Phi=0$ with (in cgs units) 
$\rho=1$, $\nu=0.1$, $\Gamma=21.5$, and $\omega=200$.  Curves are labelled by the 
solutions generated (see text) and are dashed in the case of Hopf bifurcation.  The clockwise 
circular path is used to generate the bifurcation diagram in Fig.~\ref{fig:12maxbifdiag} and is 
at a radius $\sqrt{\lambda^2+\mu^2}=0.0128$.}
\label{fig:12maxbifsets}
\end{figure}

We provide here some numerical examples using coefficients derived from the Zhang-Vi\~nals 
equations~(\ref{eq:ZV}).   Figure~\ref{fig:12maxbifsets} shows the bifurcations that occur for  
%%%%%%%%%%%%%%%%%%%%%%%%%%%%%%%%%%%%%%%%FIG
%%%%%%%%%%%%%%%%%%%%%%%%%%%%%%%%%%%%%%%%FIG
$m$:$n = 1$:2, $\Phi=0$, $\rho=1$, $\nu=0.1$, $\Gamma=21.5$, and $\omega=200$ (cgs units); 
for these parameters $\gamma=0.0341$ and we find $\alpha=-0.265$, $\beta=0.281$, 
$a=-6.136$, $b=0.0266$, $c=-0.98$, $d=-2.076$, and $e=-2.234$.    We use $m$:$n=1$:2 rather 
that $m$:$n=3$:2 (as in Section~\ref{sec:ZV}) mainly for pedagogical reasons: although the 
bifurcations seen with other forcing ratios such as 3:2 (and comparable physical parameters) are the 
same as those found in Fig.~\ref{fig:12maxbifsets} they tend to be somewhat more bunched 
together thus more difficult to illuminate.   The use of 1:2 forcing also has the advantage of the large 
${\cal O}(1/\gamma)$ oscillations with $\Phi$ in $b$, $c$, and $d$ (see Eqs.~(\ref{eq:SWcoef})); 
this leads to greater control over the normal form coefficients and hence to a wider range of 
dynamical possibilities (for example, this $\Phi$-dependence made it easier to find the 
interesting heteroclinic behavior shown in Fig.~\ref{fig:12chaotic}).

\begin{figure}[ht]
\centerline{\includegraphics{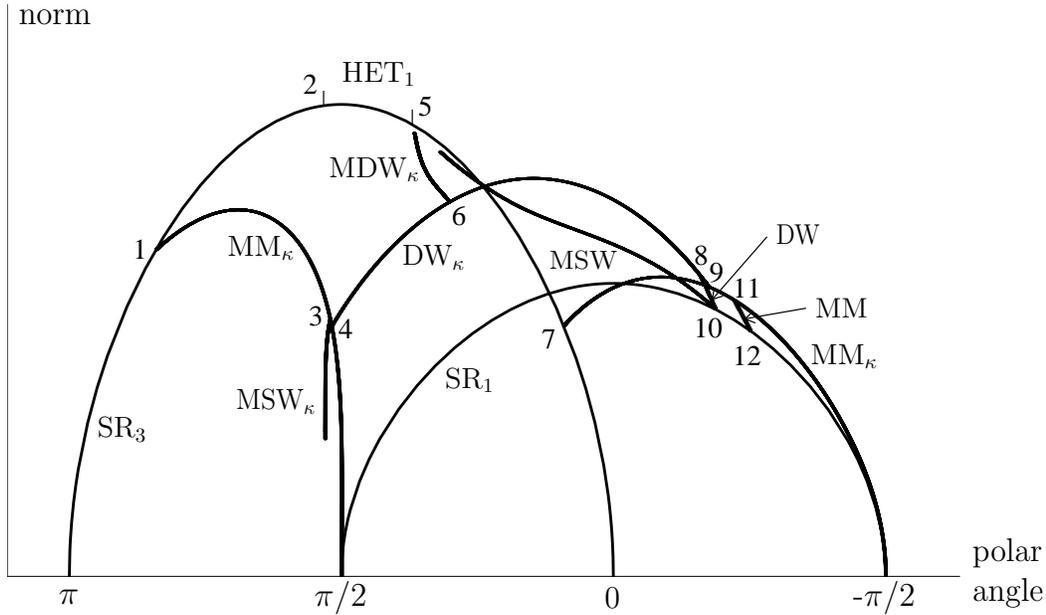}}
\caption{Bifurcation diagram for the clockwise circular path shown in 
Fig.~\ref{fig:12maxbifsets}}
\label{fig:12maxbifdiag}
\end{figure}

If one traverses the clockwise circular path shown in Fig.~\ref{fig:12maxbifsets} the resulting 
bifurcation diagram is that of Fig.~\ref{fig:12maxbifdiag}.   Initially, in the third quadrant
%%%%%%%%%%%%%%%%%%%%%%%%%%%%%%%%%%%%%%%%FIG
%%%%%%%%%%%%%%%%%%%%%%%%%%%%%%%%%%%%%%%%FIG
of the $(\lambda,\mu)$ plane, all initial conditions are attracted to the stable flat state O.  As the 
polar angle ($\phi$, say) is decreased through $\pi$ this flat state becomes unstable to 
(a circle of) standing rolls of type SR$_3$.  At $\phi \simeq 2.64$ (point 1 in Fig.~\ref{fig:12maxbifdiag}) 
these stable SR$_3$ states in turn undergo a supercritical symmetry-breaking bifurcation to (a torus of) 
$\kappa$-symmetric mixed modes MM$_\kappa$ satisfying $\theta=\pi$.  These MM$_\kappa$ states 
are stable until a subcritical Hopf bifurcation at $\phi \simeq 1.64$ (point 3) producing 
$\kappa$-symmetric modulated standing waves MSW$_\kappa$.  The MSW$_\kappa$ are 
themselves destroyed in a heteroclinic bifurcation (point 2, marked by a small dash on the SR$_3$ 
branch) joining O and SR$_3$ (see \cite{PorKno01}) at $\phi \simeq 1.67$.   The same global bifurcation 
marks the birth of structurally (and asymptotically) stable heteroclinic cycles (HET$_1$) connecting 
points on the circle of SR$_3$ states to their ``$\pi$ translates".   After the Hopf bifurcation renders 
MM$_\kappa$ unstable the heteroclinic cycles HET$_1$ are the only attractors of the 
system~(\ref{eq:SW1}-\ref{eq:SW3}).    These HET$_1$ remain stable until $\phi \simeq 1.16$ 
(point 5; this is when $\mu=\lambda e/c$ and the two eigenvalues of Eq.~(\ref{eq:SR3eig}), one 
positive and one negative, are of equal magnitude).  The bifurcation that occurs here simultaneously 
produces stable MDW$_\kappa$ and unstable MSW (the fact that the MSW branch does not visibly 
originate here is due to numerical limitations).  The stable MTW$_\kappa$ branch describes modulations 
of ever decreasing amplitude as it approaches the branch of DW$_\kappa$ states, themselves present 
since an ${\cal R}$ symmetry-breaking bifurcation on MM$_\kappa$ at $\phi \simeq 1.63$ (point 4), 
and is destroyed in a Hopf bifurcation at $\phi \simeq 0.95$ (point 6).  The DW$_\kappa$ are then stable 
until a supercritical $\kappa$ symmetry-breaking bifurcation to DW at $\phi \simeq -0.5$ (point 8); 
this occurs just prior to their termination at $\phi \simeq -0.55$ (point 9) on the {\it other} 
MM$_\kappa$ branch, i.e., the one with $\theta=0$ existing between $\phi = -\pi/2$ and a bifurcation 
from SR$_3$ at $\phi \simeq 0.29$ (point 7).  The stable DW branch terminates, along with the 
unstable MSW branch, in a Hopf bifurcation on SR$_1$ at $\phi \simeq -0.59$ (point 10).  The 
SR$_1$ rolls are stable between this Hopf bifurcation and a subcritical bifurcation to MM at 
$\phi \simeq -0.79$ (point 12).  This MM branch also connects to (and stabilizes) the 
MM$_\kappa$ branch at $\phi \simeq -0.7$ (point 11).   Finally, the MM$_\kappa$ states (with 
$\theta=0$) are stable between this last bifurcation and $\phi=-\pi/2$ where they vanish, along with 
SR$_1$, restoring stability to the flat state.

\begin{figure}[ht]
\centerline{\includegraphics[width=3.5in]{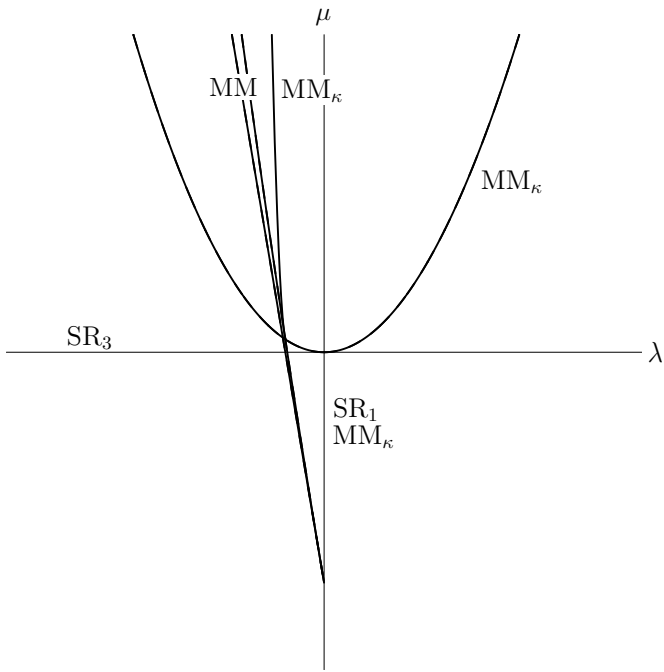}}
\caption{Bifurcation sets for $m$:$n = 1$:2 with the same parameters as Fig.~\ref{fig:12maxbifsets} 
except that $\Phi=\pi/2$.  For this choice of phase $\alpha\beta > 0$ and there are no DW$_\kappa$, 
MDW$_\kappa$, DW, MSW$_\kappa$, MSW, or heteroclinic solutions.}
\label{fig:12minbifsets}
\end{figure}

If, in contrast to the case of Fig.~\ref{fig:12maxbifsets}, one adjusts the forcing phases so that $\Phi=\pi/2$ 
then the (leading order) expressions for $\alpha$ and $\beta$ of Eqs.~(\ref{eq:SWcoef}) vanish.  With the 
same physical parameters as Fig.~\ref{fig:12maxbifsets} it turns out that the first nonvanishing contributions 
have the {\it same} algebraic sign, i.e., $\alpha \beta > 0$.   Specifically, we find $\alpha=0.0784$, 
$\beta=0.0744$, $a=0.549$, $b=-0.523$, $c=-0.0155$, $d=-3.266$, and $e=-2.227$.   The fact that 
$\alpha \beta > 0$ explains the dramatically different (and less rich) unfolding shown in 
Fig.~\ref{fig:12minbifsets}.  In this case  
%%%%%%%%%%%%%%%%%%%%%%%%%%%%%%%%%%%%%%%%FIG
%%%%%%%%%%%%%%%%%%%%%%%%%%%%%%%%%%%%%%%%FIG
there are no drifting solutions (DW$_\kappa$, MDW$_\kappa$, DW) or heteroclinic cycles (HET$_1$, 
HET$_2$).   Furthermore, there are no modulated standing waves MSW$_\kappa$ or MSW 
(in was shown in \cite{Dan86} that $\alpha\beta<0$ is necessary for MSW$_\kappa$).  A second, 
somewhat surprising, feature of the $\Phi=\pi/2$ coefficients is the large change in the 
self-interaction coefficient $a$: it is now positive whereas at $\Phi=0$ it was $-6.136$.  
Such substantial $\Phi$-dependence is not an ingredient of Eqs.~(\ref{eq:SWcoef}) and is a 
result of ``1:2" linear resonances, i.e., harmonic modes satisfying $\bs{k}=2\bs{k}_1$ or 
$\bs{k}=2\bs{k}_2$.  The influence of these damped modes (see \cite{TopSil02}) is especially 
strong here because $k_3 \approx 1.86$ is quite close to $2k_1 \approx 1.94$ (the damping of 
the $2k_1$ modes is thus very small and they nearly satisfy the 1:2 temporal resonance 
condition $m = \Omega(2k_1)$; in this case $m=n/2$).  For comparison, note that the $2k_3$ 
modes are not ``almost" neutral and the corresponding self-interaction coefficient $e$ shows no 
significant dependence on $\Phi$, in agreement with Eqs.~(\ref{eq:SWcoef}).   Because $a>0$ 
the bifurcation of SR$_1$ and MM$_\kappa$ from the flat state is now subcritical (observe how 
the MM bifurcations from SR$_1$ and MM$_\kappa$ have switched to the left side of 
$\lambda=0$ in Fig.~\ref{fig:12minbifsets}) and the cubic truncation~(\ref{eq:SW1}-\ref{eq:SW3}) 
is no longer well-behaved (some trajectories  diverge).

%%%%%%%%%%%%%%%%%%%%%%%%%%%%%%%%%%%%%%%%%%%%%%%
%%%%%%%%%%%%%%%%%%%%%%%%%%%%%%%%%%%%%%%%%%%%%%%
\section{Conclusions}
\label{sec:conclusion}

The weakly nonlinear dynamics of a class of resonant triads occurring in the vicinity of the bicritical 
point of the two-frequency forced Faraday problem has been studied in detail.  The approach 
taken for this resonant triad can be easily extended to a wide range of patterns produced by 
multi-frequency forcing (see \cite{PorSil02}), and is based on very general symmetry considerations.  
In particular, we make use of the fact that the weakly damped problem possesses additional structure 
beyond the more obvious spatial and discrete temporal symmetries, i.e., it must satisfy the constraints 
of continuous time translation and time reversal (when these are viewed as parameter symmetries) 
and respect the nearby Hamiltonian structure of the undamped problem.   The effect of these ``broken" 
symmetries is straightforward to establish in the context of TW equations~(\ref{eq:TWf1}-\ref{eq:TWf3}), 
and in the reduced SW equations~(\ref{eq:SW1}-\ref{eq:SW3}) reveals itself in the structure of the 
normal form coefficients~(\ref{eq:SWcoef}).  Most notably, we found that the SW coefficients of the 
quadratic resonance terms ($\alpha$ and $\beta$) could be expected to be of opposite sign, exhibit simple 
harmonic dependence on the phase $\Phi$, and scale as the damping $\gamma$ to the 
power $(m+n-3)/2$.  Each of these predictions was supported in the direct numerical calculation 
of coefficients from the Zhang-Vi\~nals equations~(\ref{eq:ZV}).  Moreover, the exponential 
decrease in the size of the resonance terms with $m+n$ provides an explanation for the lack of experimental 
observations \cite{ArbFin02,ArbFin00} of this triad with large values of $m$ and $n$.   Recall, however, that this 
simple prediction (as well as the other scaling laws) emerged after we expanded the SW normal form 
coefficients in powers of $f_m$, $f_n$, and $\gamma$ (see Eqs.~(\ref{eq:upsilon}-\ref{eq:coef2})).  
Particularly in the case of $\gamma$ it is not clear that the assumption of analyticity is a good one -- the 
addition of damping is known to act as a singular perturbation (see, e.g., \cite{Phi77,MarKno97}).  For the 
quasipotential equations of Zhang and Vi\~nals the numerical agreement (and hence the justification for 
a simple Taylor expansion) is quite convincing, but we speculate that a more rigorous treatment using 
the Navier-Stokes equations with realistic boundary conditions and finite fluid depth may lead to a 
departure, in some regimes, from the scalings given here.  This departure will presumably be small in 
weakly damped systems of large depth where the Zhang-Vi\~nals equations are expected to be valid.

The resonant triad system~(\ref{eq:SW1}-\ref{eq:SW3}) exhibits extremely rich dynamics, 
much more so in the case where the resonant coefficients satisfy $\alpha \beta<0$.  That the 
underlying Hamiltonian structure makes this the typical case for weak damping and, furthermore, 
that this situation depends on the relative phase of the two forcing terms, is a result we have 
emphasized throughout this paper.  When $\alpha\beta<0$ there exist several types of  modulated 
standing waves, drifting waves (of both steady and modulated character), and heteroclinic cycles 
with associated phenomena like the chaotic attractor of Fig.~\ref{fig:12chaotic}. It must be 
admitted, however, that none of these more complicated solutions has yet been reported in the 
experimental literature; other than simple roll states, only the mixed mode MM$_\kappa$ has
been established experimentally~\cite{ArbFin00}.  It is quite possible that the remaining solutions 
are simply unstable when all types of perturbations are considered (the system may prefer hexagons 
or squares, for example).  It might also be that the use of finite containers plays a critical role
 in preventing many of these solutions from appearing (the drifting waves DW$_\kappa$, 
DW, and MTW$_\kappa$ obviously cannot exist as such in nonperiodic bounded domains).  It 
would be interesting to see if a more deliberate search, perhaps in a very large aspect ratio 
container, would reveal more features of the dynamics presented in Section~\ref{sec:dyn}.  On the 
other hand, it would miss the point to focus too narrowly on the resonant triad treated here.   This 
paper is also intended to serve as a detailed illustration of a very general method with wide applicability 
to parametrically forced problems, particularly those with forcing functions composed of multiple 
frequencies -- for example, in \cite{PorSil02} we considered two-frequency forced hexagons 
and two-mode superlattices \cite{ArbFin98}.  The symmetry-based approach of this paper provides, for 
weakly damped systems, a means of determining scaling laws and other important qualitative features 
of the normal form coefficients without resorting to a full center manifold reduction from the 
governing equations (often an exceedingly difficult task).  The results also suggest how one might 
begin to think about designing forcing functions that enhance, suppress, or otherwise control the 
patterns of interest (superlattices, for example).  This very intriguing direction will be pursued in 
future work \cite{PorTopSil}.

\begin{ack}
We thank J. Fineberg, M. Golubitsky, E. Knobloch, H. Riecke, P. Umbanhowar, and in particular 
C.M. Topaz for very helpful discussions.  This work was supported by NASA Grant No. NAG3-2364 
and NSF Grant No. DMS-9972059.
\end{ack}

%%%%%%%%%%%%%%%%%%%%%%%%%%%%%%%%%%%%%%%%%%%%%%%%%
\appendix
\section{Appendix}

We give here the expressions for $J_1, ..., J_7$ appearing in Eqs.~(\ref{eq:SWcoef}).
\begin{align}
&J_1 =\frac{2\varrho_r^2}{\tilde{\mu}_i^2}\big(\hat{\upsilon}_n+\frac{\upsilon_r
\tilde{\lambda}_n}{\tilde{\lambda}_i}-\upsilon_n C_1\big), \quad \;
{\rm where}\quad 
C_1 = \frac{\upsilon_\gamma}{\upsilon_r} + \frac{\upsilon_m\upsilon_r}{\tilde{\lambda}_i^2} + 
\frac{\upsilon_n\varrho_r^2}{\upsilon_r\tilde{\mu}_i^2},\label{eq:J1}\\
&J_2 =\frac{2\upsilon_r^2}{\tilde{\lambda}_i^2}\big(\hat{\varrho}_m+\frac{\varrho_r
\tilde{\mu}_m}{\tilde{\mu}_i}-\varrho_n C_2\big),\quad 
{\rm where}\quad
C_2 = \frac{\varrho_\gamma}{\varrho_r}+\frac{\varrho_m\upsilon_r^2}{\varrho_r\tilde{\lambda}_i^2}+
\frac{\varrho_n\varrho_r}{\tilde{\mu}_i^2},\label{eq:J2}\\
&J_3={\rm a}_r+{\rm b}_r-C_1({\rm a}_i+{\rm b}_i)+\frac{\upsilon_r}{\tilde{\lambda}_i}
\big(({\rm r}_2)_i-({\rm q}_1)_i-({\rm q}_2)_i \big),\label{eq:J3}\\
&J_4=\delta_r+{\rm c}_r+{\rm d}_r-C_1(\delta_i+{\rm c}_i+{\rm d}_i)+
\frac{\upsilon_r}{\tilde{\lambda}_i}\big(({\rm r}_3)_i-({\rm r}_1)_i-({\rm q}_3)_i
-({\rm q}_4)_i\big),\label{eq:J4}\\
&J_5={\rm e}_r+{\rm f}_r-C_1({\rm e}_i+{\rm f}_i)-\frac{\upsilon_r}{\tilde{\lambda}_i}
\big(({\rm q}_5)_i+({\rm q}_6)_i\big)+\frac{\varrho_r}{\tilde{\mu}_i}
\big(({\rm r}_5)_i-({\rm r}_4)_i\big),\label{eq:J5}\\
&J_6={\rm g}_r+{\rm h}_r-C_2({\rm g}_i+{\rm h}_i)-\frac{\varrho_r}{\tilde{\mu}_i}
\big(({\rm s}_1)_i+({\rm s}_2)_i\big)+\frac{\upsilon_r}{\tilde{\lambda}_i}
\big(({\rm u}_2)_i-({\rm u}_1)_i\big),\label{eq:J6}\\
&J_7={\rm l}_r+{\rm p}_r-C_2({\rm l}_i+{\rm p}_i)+\frac{\varrho_r}{\tilde{\mu}_i}
\big(({\rm u}_3)_i-({\rm s}_3)_i-({\rm s}_4)_i\big).\label{eq:J7}
\end{align}

\bibliography{rt}

\end{document}